\documentclass[12pt,letterpaper]{article}
\pdfoutput=1
\usepackage{jheppub}
\usepackage{amsfonts, amsthm}
\usepackage[english]{babel}
\usepackage[utf8]{inputenc}
\usepackage{slashed}
\usepackage{mathrsfs}
\usepackage{amssymb}
\usepackage{color}
\hypersetup{unicode}
\usepackage{natbib}

\newcommand{\eq}{\begin{equation}}
\newcommand{\feq}{\end{equation}}
\newcommand{\eqn}{\begin{eqnarray}}
\newcommand{\feqn}{\end{eqnarray}}

\newcommand{\ma}[1]{\mbox{$\mathcal{#1}$}}

\newcommand{\mrm}[1]{\mbox{$\mathrm{#1}$}}
\newcommand{\D}{{\rm d}}

\title{Surface defects in the D4-D8 brane system}

\author[a]{Giuseppe Dibitetto}
\author[b]{Nicol\`o Petri}

\affiliation[a]{Institutionen f\"or fysik och astronomi, University of Uppsala,  Box 803, SE-751 08 Uppsala, Sweden.}
\affiliation[b]{Department of Mathematics, Bo\u{g}azi\c{c}i University,
34342, Bebek, Istanbul, Turkey.}

\emailAdd{giuseppe.dibitetto@physics.uu.se}
\emailAdd{nicolo.petri@boun.edu.tr}

\abstract{A new class of exact supersymmetric solutions is derived within minimal $d=6$ $F(4)$ gauged supergravity. These flows are all characterized by a non-trivial radial profile for the 2-form gauge potential included into the supergravity multiplet. In particular three solutions within this class are featured by an $\mrm{AdS}_3$ foliation of the 6d background and by an $\mrm{AdS}_6$ asymptotic geometry. Secondly, considering the simplest example of these, we give its massive IIA uplift describing a warped solution of the type $\mrm{AdS}_3\times S^2\times S^3$ fibered over two intervals $I_r\times I_\xi$. We interpret this background as the near-horizon of a D4-D8 system on which a bound state D2-NS5-D6 ends producing a surface defect. Finally we discuss its holographic dual interpretation in terms of a $\ma N=(0,4)$ $\mrm{SCFT}_2$ defect theory within the $\ma N=2$ $\mrm{SCFT}_5$ dual to the $\mrm{AdS}_6\times S^4$ massive IIA warped vacuum.}


\begin{document}
\maketitle
\flushbottom

\section{Introduction}

The most peculiar feature of the quantum string theory spectrum is the presence of extended objects of non-perturbative nature, which are referred to as branes.
Therefore, branes as such are the key to the non-perturbative aspects of string theory. Even if a lot of progress has been made in this respect, all main insights in this direction are still coming from the low-energy description of brane systems. For this reason, the search for new supersymmetric solutions within supergravity theories, as well as engineering novel examples of SCFTs emerging from branes should be considered as the most practical, concrete and predictive playgrounds for producing quantitative results concerning the physics of strings propagating within ten dimensional spacetime.


The aim of this paper is to take some further steps in this direction by considering the holographic realization of defect conformal field theories arising from brane systems. Generally speaking, these are CFTs defined on a defect hypersurface within the background of a higher-dimensional bulk CFT \cite{Cardy:1984bb,Cardy:1989ir,Cardy:1991tv,McAvity:1993ue,McAvity:1995zd,Behrend:1999bn}. From the point of view of this ``mother'' theory, the presence of the defect is realized through a deformation associated to a position-dependent coupling. This deformation turns out to partially break conformal invariance in the bulk, while only preserving the conformal transformations leaving the defect CFT intact. As an immediate consequence, the one-point correlation functions are no longer vanishing, and a non-trivial displacement operator appears. This a sign of the fact that the energy-momentum tensor needs not be conserved in the presence of the defect.

The first realizations of defect CFTs in string theory were constructed in \cite{Karch:2000gx}. Then many other examples and applications followed (for a non-exhaustive list of references on conformal defects in string theory and holography see \cite{DeWolfe:2001pq,Bachas:2001vj,Erdmenger:2002ex,Constable:2002xt,Aharony:2003qf,Bak:2003jk,Clark:2004sb,Kapustin:2005py,Clark:2005te,DHoker:2006qeo,DHoker:2006vfr,Buchbinder:2007ar,DHoker:2007zhm,DHoker:2007hhe,Lunin:2007ab,Gaiotto:2008sa,Gaiotto:2008sd,Aharony:2011yc,Chiodaroli:2010ur,Chiodaroli:2011fn,Chiodaroli:2012vc,Gutperle:2012hy,deLeeuw:2015hxa,Billo:2016cpy,Karndumri:2017bqi,Dibitetto:2017tve,Dibitetto:2017klx,DelZotto:2018tcj,Karndumri:2018yiz}). The key idea is to let defect CFTs emerge from some particular supersymmetric brane configurations in which some ``defect branes'' end on a given brane system, which is known to give rise to an AdS vacuum in the near-horizon limit. The main effect of these intersections is to break partially the isometry group of the AdS vacuum of the original brane system and to produce a lower-dimensional warped AdS solution. The defect CFT describes the boundary conditions defining the intersection with the defect branes and the warping of the corresponding background describes the backreaction of the defect onto the bulk geometry. This may be viewed as the supergravity picture associated to the position-dependent deformation of the ``mother'' SCFT, dual to the original higher-dimensional AdS vacuum.

More concretely, let us consider a SUSY $\mathrm{AdS}_d$ closed string vacuum associated with the near-horizon of some brane system, where we furthermore assume the existence of a consistent truncation linking the 10d (or 11d) picture to a solution in a $d$-dimensional gauged supergravity describing the excitations around the $\mrm{AdS}_d$ vacuum.
If some defect branes end on this system, then we have a bound state with a $(p+1)$-dimensional worldvolume whose physics  is captured by a $d$-dimensional Janus-type background
\begin{equation}
 ds_d^2=e^{2U(r)}\,ds^2_{{\scriptsize \mrm{AdS}_{p+2}}}+e^{2V(r)}\,dr^2+e^{2W(r)}\,ds^2_{d-p-3}\,.
 \label{slicing} 
\end{equation} 
 The $d$-dimensional background is thus characterized by a $\mrm{AdS}_{p+2}$ slicing and an asymptotic region locally described by the $\mrm{AdS}_d$ vacuum\footnote{We point out that the main difference between this case and the one of RG flows across dimensions can be observed by considering the ``radial'' coordinates giving rise to the AdS vacua respectively in the UV and IR. In a conformal defect, the radii of the $\mrm{AdS}_{p+2}$ and $\mrm{AdS}_d$ are different, while in a supergravity solution describing an RG flow across dimensions, the AdS backrounds arising in the UV and in the IR are described by the same radial coordinate. Conformal defects and RG flows across dimensions are somehow two complementary descriptions. For example one may guess the existence of more general flows involving $r$ as well as the radial coordinate of $\mrm{AdS}_{p+2}$ describing a geometry where the metric \eqref{slicing} is replaced by an $\mathbb{R}^{1,p}$ slicing of the $d$-dimensional background.}. The solutions like \eqref{slicing} can be then consistently uplifted producing warped geometries of the type $\mrm{AdS}_{p+2}\times \ma {M}_{d-p-2}\times \Sigma_{D-d}$, where $\ma {M}_{d-p-2}$ is realized as a fibration of the $(d-p-3)$-dimensional transverse manifold over the interval $I_r$ and $\Sigma_{D-d}$ is the internal manifold of the truncation with $D=10 \text{ or }11$. From the point of view of the dual field theories, this is exactly the supergravity realization of a defect $\mrm{SCFT}_{p+1}$ within the ``mother'' $\mrm{SCFT}_{d-1}$.

In this paper we consider D4-D8 systems in massive IIA string theory and its intersection with D2-NS5-D6 defect branes. It is well-known that stacks of coincident D4 branes localized on D8 branes and in the presence of O8 planes are described at the horizon by a warped vacuum $\mathrm{AdS}_6\times S^4$ \cite{Brandhuber:1999np}. The dual picture of this vacuum is realized by a matter-coupled $\ma N=2$ $\mrm{SCFT}_5$ arising as a fixed point of the 5d quantum field theory living on the worldvolume of the D4s \cite{Intriligator:1997pq,Seiberg:1996bd,Ferrara:1998gv,Brandhuber:1999np}. For a non-exhaustive list of references on $\mrm{AdS}_6$ vacua in string theory and $\mrm{AdS}_6/\mrm{CFT}_5$ correspondence we will refer to \cite{DeWolfe:1999hj,Jafferis:2012iv,Assel:2012nf,Bergman:2012qh,Bergman:2012kr,Passias:2012vp,Lozano:2012au,Karndumri:2012vh,Bergman:2013koa,Lozano:2013oma,Apruzzi:2014qva,Pini:2014bea,Kim:2015hya,Karndumri:2016ruc,DHoker:2016ujz,DHoker:2016ysh,Passias:2018swc,Bergman:2018hin,Bah:2018lyv}.

Massive IIA string theory can be consistently truncated around the $\mathrm{AdS}_6\times S^4$ vacuum \cite{Cvetic:1999un} and the theory produced by this truncation is $d=6$, $\ma N=(1,1)$ gauged supergravity, also known as $F(4)$ gauged supergravity \cite{Romans:1985tw}. The minimal incarnation of this theory will be the main tool of this paper and, within this context, we will be able to derive a new class of analytic BPS solutions characterized by a running profile for the 2-form gauge potential included into the supergravity multiplet. This new class of flows will be presented by starting from the simplest 6d background compatible with the presence of the 2-form, to subsequently move to more complicated 6d geometries. The main results are thus represented by three backgrounds of the type \eqref{slicing}, namely, warped solutions of the type $\mrm{AdS}_{3}\times \ma {M}_{3}$ admitting a locally $\mrm{AdS}_6$ asymptotic geometry with a 2-form charge. In particular one of these backgrounds is non-singular in the IR and, in this limit, the geometry is locally given by $\mrm{AdS}_3\times T^3$.

Among these warped $\mrm{AdS}_{3}\times \ma {M}_{3}$ solutions, we then consider the simplest one, given by a ``charged'' domain wall with a running profile for the 2-form and we interpret the singular behavior appearing in the IR regime as a brane singularity associated to D2-NS5-D6 defect branes ending on the D4-D8 system. The key point of this interpretation is based on the presence of the 2-form that turns out to be the related to the $F_{(4)}$, $F_{(2)}$ and $H_{(3)}$ fluxes in the 10d picture.
Thanks to the uplift formula, we then obtain the corresponding 10d backround written as $\mrm{AdS}_{3}\times S^2\times S^3$ fibered over two intervals $I_r\times I_\xi$, where $\ma M_3$ is realized by an $S^2$ fibration over $I_r$ and the 4d squashed sphere defining the truncation is written as an $S^3$ fibration over $I_\xi$. Then we discuss the relations of the 10d background $\mrm{AdS}_{3}\times S^2\times S^3\times I_r\times I_\xi$ with the near-horizon geometry of the brane intersection D2-D4-NS5-D6-D8 found in \cite{Dibitetto:2017klx} and we formulate the holographic interpretation of the 6d charged domain wall as a defect $\ma N=(0,4)$ $\mrm{SCFT}_2$ within the $\ma N=2$ $\mrm{SCFT}_5$. Finally we test this interpretation by deriving the one-point functions of the defect both from holographic arguments and conformal perturbation expansion, and we find agreement in the position-dependence for the coupling driving the deformation produced by the defect.

\section{The D4-D8 System and $\mrm{AdS}_6/\mrm{CFT}_5$}
\label{D4D8system}

Let us consider the brane system discussed in \cite{Seiberg:1996bd,Brandhuber:1999np,Ferrara:1998gv}. The construction starts from a probe five-brane brane in type I string theory on $\mathbb{R}^9\times S^1$ whose worldvolume is wrapping the circle. Performing a T-duality along the circle we obtain a four-brane in type I' on the interval $S^1/\mathbb{Z}_2$ with two $\mathrm{O}8$ planes in the fixed points. Then the four-brane can be interpreted as a D4 brane in massive IIA string theory located at a point of the interval. In order to cancel the $-16$ charge units carried by the O8 planes, one has to include at least $16$ D8 branes whose position is described by the moduli appeared after dualizing. Then a slightly more general construction involving two D8 stacks can be considered, one of each consisting of $N_f$ and $16-N_f$ D8 branes, respectively. 

Let us now move to discussing the worldvolume theory of this construction. Along the interval $S^1/\mathbb{Z}_2$, the gauge group of the theory on the D4 brane is broken to $\textrm{U}(1)$, but at the two endpoints a larger gauge symmetry is restored. In particular, if the D4 and $N_f$ D8 branes are located at one orientifold and the other $16-N_f$ at the other O8, then we have a $d=5$ $\ma N=2$ Yang-Mills theory with gauge group $\textrm{SU}(2)$. The 5d vector multiplet includes a gauge field and a real scalar describing the locus of the D4 along $S^1/\mathbb{Z}_2$. The matter content is given by $N_f$ hypermultiplets in the fundamental, arising from open strings streched between the D4 and the D8 branes, and by an antisymmetric massless hypermultiplet coming from the D4 brane.
The supercharges and the scalars coming from the antisymmetric hypermultiplet transform as a doublet under the R-symmetry group, that is given by $\textrm{SU}(2)_R$. The global symmetry of the theory is $\textrm{SU}(2)\times \textrm{SO}(2N_f)\times \textrm{U}(1)_I$, where the $\textrm{SU}(2)$ factor is associated to the antisymmetric hyper, the $\textrm{SO}(2N_f)$ one is related to the $N_f$ hypers in the fundamental and finally the extra $\textrm{U}(1)_I$ corresponds to the instanton number conservation\footnote{It is related to the 5d conserved current $\star_{\,5}(F\wedge F)$ \cite{Seiberg:1996bd,Ferrara:1998gv}.}.

The above construction can be extended to a stack of $N$ coinciding D4 branes entirely localized on the $N_f$ D8 branes at a 9-dimensional orientifold and other $16-N_f$ D8 branes at the other O8 plane. In this case we have a $\ma N=2$ SYM theory with gauge group $\mrm{USp}(N)$ coupled to $N_f$ ``quark'' hypers and to an antisymmetric hyper.

If the number of flavors is such that $N_f<8$, the theory introduced above has a non-trivial fixed point at the origin of the Coulomb branch, given by $\mathbb{R}^+$ and the global symmetry associated to the Higgs branch is then enhanced to $\textrm{SU}(2)\times \textrm{E}_{N_f+1}$ \cite{Seiberg:1996bd}. This fixed point is described by a $\ma N=2$ $\mrm{SCFT}_5$ with $\mrm{USp}(N)$ gauge group and couplings to matter given by $N_f$ fundamental and one antysimmetric hypermultiplets.

The low-energy description of the above brane system is naturally realized in massive IIA supergravity\footnote{See Appendix \ref{app:massiveIIA} for a brief review on massive IIA supergravity.}. It turns out that this construction includes an $\mrm{AdS}_6$ vacuum in its near-horizon limit and this corrisponds to a fixed point in the RG flow of the 5d worldvolume theory of the D4 branes \cite{Brandhuber:1999np,Ferrara:1998gv}.
\begin{table}[h!]
\renewcommand{\arraystretch}{1}
\begin{center}
\scalebox{1}[1]{
\begin{tabular}{c||c c c c c || c c||c c c}
branes & $t$ & $y^{1}$ & $y^{2}$ & $y^{3}$ & $y^{4}$ & $z$ & $\rho$ & $\theta^{1}$ & $\theta^{2}$ & $\theta^{3}$ \\
\hline 
D8 & $\times$ & $\times$ & $\times$ & $\times$ & $\times$ & $-$ & $\times$ & $\times$ & $\times$ & $\times$ \\
D4 & $\times$ & $\times$ & $\times$ & $\times$ & $\times$ & $-$ & $-$ & $-$ & $-$ & $-$ 
\end{tabular}
}
\end{center}
\caption{{\it The brane picture underlying the 5d $\ma N=2$ SCFT defined by the D4-D8 system. The system is $\text{BPS}/4$ and
the $\textrm{AdS}_{6}\times S^4$ vacuum is realized by a combination of $\rho$ and $z$.}} \label{Table:BO}
\end{table}
Let us now consider the supergravity solution describing the simplest realization of such D4-D8 system. Given a D4 probing a D8 background with worldvolume along the coordinates $(t, x^1, x^2, x^3, x^4)$ and located at points $(z, \rho, \theta^1, \theta^2, \theta^3)$, the massive IIA field configuration has the following form \cite{Brandhuber:1999np,Youm:1999zs,Imamura:2001cr}
  \begin{equation}
   \begin{split}
    d s_{10}^2&=H_{\mathrm{D}8}^{-1/2}\,H_{\mathrm{D}4}^{-1/2}\,d s^2_{\mathbb{R}^{1,4}}+H_{\mathrm{D}8}^{1/2}\,H_{\mathrm{D}4}^{1/2}\,dz^2+H_{\mathrm{D}8}^{-1/2}\,H_{\mathrm{D}4}^{1/2}\,\left(d \rho^2+\rho^2\,ds^2_{S^3} \right)\ ,\\
   e^{\Phi}&=g_s\,H_{\mathrm{D}8}^{-5/4}\,H_{\mathrm{D}4}^{-1/4}\ ,\qquad C_{(5)}=\frac{1}{g_s\,H_{\mathrm{D}4}}\ ,
   \label{BOsolution}
   \end{split}
  \end{equation}
where $H_{\mathrm{D}8}=H_{\mathrm{D}8}(z)$ and $H_{\mathrm{D}4}=H_{\mathrm{D}4}(z,\rho)$ are suitable functions given by
\begin{equation}
H_{\mathrm{D}4}(z, \rho)=1+\frac{Q_{{\scriptsize{\mrm{D}4}}}}{(\rho^2+\frac{9}{4}\,g_sm\,z^3)^{5/3}}\qquad \text{and}\qquad H_{\mathrm{D}8}(z)=g_s\,m\,z\ ,
\label{BOsolution2}
\end{equation}
while $ds^2_{S^3}$ is the metric on the round $S^3$ parametrized by the coordinates $\theta^i$. This solution depends on two parameters, respectively given by the D4 charge $Q_{{\scriptsize{\mrm{D}4}}}$, and the D8 charge $Q_{{\scriptsize{\mrm{D}8}}}\,=\,g_s m$, $m$ being the Romans' mass.
The background \eqref{BOsolution} satifies the 10d equations of motion \eqref{massiveIIAeoms}, while the Bianchi identities \eqref{massiveIIAbianchi} are trivially satified. This last feature may be viewed as a consequence of the fact that the Hanany-Witten effect does not occur in D4-D8 constructions.

The AdS geometry arising in the near-horizon limit can be understood by introducing the following change of coordinates
\begin{equation}
 \rho=\zeta\,\cos\alpha \qquad \text{and}\qquad z=\left(\frac{3}{2}\right)^{2/3}g_s^{-1/3}m^{-1/3}\,\zeta^{2/3}\,\sin^{2/3}\alpha\ ,
\end{equation}
the functions \eqref{BOsolution2} take the following form
\begin{equation}
 H_{\mathrm{D}8}=\left(\frac{3}{2}\right)^{2/3}g_s^{2/3}m^{2/3}s^{2/3}\zeta^{2/3}\qquad \text{and}\qquad H_{\mathrm{D}4}=1+\frac{Q_{{\scriptsize{\mrm{D}4}}}}{\zeta^{10/3}}\ ,
\end{equation}
with $s=\sin\alpha$ and $c=\cos\alpha$.

In this new coordinate system, the near-horizon limit is given by $\zeta \rightarrow 0$ and it corresponds to the regime in which the ``$1$'' in $H_{\mathrm{D}4}(\zeta)$ can be dropped. In this case the metric in \eqref{BOsolution} can be cast in the following form \cite{Brandhuber:1999np}
\begin{equation}
 \begin{split}
  d s^2_{10}&= \left(\frac32 g_s m\, s \right)^{-1/3}\,\left[Q_{{\scriptsize{\mrm{D}4}}}^{-1/2}ds^2_{{\scriptsize \mrm{AdS}_6}}+ Q_{{\scriptsize{\mrm{D}4}}}^{1/2} ds_{S^4}^2   \right]\ ,\\[2mm]
  ds^2_{{\scriptsize \mrm{AdS}_6}} &=\frac{9\,Q_{{\scriptsize{\mrm{D}4}}}}{4}\,\frac{du^2}{u^2}+u^2 ds^2_{\mathbb{R}^{1,4}}\ ,\\[1mm]
  ds^2_{S^4}&=d\alpha^2+c^2\,d \Omega^2_3\ ,
  \label{BOvacuum}
 \end{split}
\end{equation}
where $u=\zeta^{2/3}$. From \eqref{BOvacuum} we conclude that the near-horizon limit of \eqref{BOsolution} is described by a warped vacuum of the type $\mrm{AdS}_6\times S^4$ where $S^4$ is only the upper emisphere of a (round) 4-sphere \cite{Brandhuber:1999np}. The boundary of $S^4$ is located at $z=0$ (or at  $\alpha \rightarrow 0$) and it describes the location of the O8 plane. The isometry group of this vacuum is given by $\textrm{SO}(2,5)\times \textrm{SU}(2)\times \textrm{SU}(2)$.

If we now consider the more general case of a stack of $N=Q_{{\scriptsize{\mrm{D}4}}}$ coinciding D4 branes entirely localized on the $N_f$ D8 branes at a 9-dimensional orientifold and other $16-N_f$ D8 branes at the other O8 plane, we may conclude, following the usual holographic dictionary, that the low-energy limit of the above D4-D8 construction
enjoys two dual descriptions appearing at the near-horizon of the corresponding brane solution. In particular it turns out that massive IIA string theory on the $\mrm{AdS}_6\times S^4$ vacuum \eqref{BOvacuum} is dual to the $\ma N=2$ $\mrm{SCFT}_5$ emerging at the horizon as a fixed point of the worldvolume theory of the underlying D4-D8 system \cite{Brandhuber:1999np,Ferrara:1998gv}. In particular the two $\textrm{SU}(2)$ isometry groups of the supergravity vacuum respectively correspond to the R-symmetry group of the $\mrm{SCFT}_5$ and to the global symmetry of the antysimmetric hypermultiplet. Moreover this theory realizes the exceptional superconformal algebra $F(4)$, whose R-symmetry only includes a single $\textrm{SU}(2)_R$. As far as the number of flavors is concerned, it must satisfy $N_f<8$, and it is associated to the Romans' mass through $m=8-N_f>0$. The further enhancement to $\textrm{E}_{N_f+1}$ which is expected at the fixed point from a field-theoretical viewpoint, may be obtained in this context by observing that the dilaton blows up as $\alpha \rightarrow 0$, thus rendering the corresponding type I' string theory descriprion strongly coupled. The aforementioned enhancement can then be explained in terms of D0 brane instanton effects. These appear at the boundary and take the new gauge degrees of freedoms into account \cite{Seiberg:1996bd,Brandhuber:1999np,Ferrara:1998gv}.

\subsection{Including a NUT Charge}

In the previous subsection we reviewed the simple original construction of D4-D8 systems and the associated 5d fixed points. 
As already explained, these theories realize the exceptional superconformal algebra $F(4)$, whose R-symmetry only includes a single $\textrm{SU}(2)$ factor.
Note that the $\textrm{SU}(2)^{2}$ isometries of the background in \eqref{BOvacuum} can be broken to $\textrm{SU}(2)$ by writing the round $S^{3}$ metric 
as a Hopf fibration of $S^{2}$ over $S^{1}$, \emph{i.e.}
\begin{equation}
d s^2_{S^3} \ = \ \frac{1}{4}\,d s^2_{S^2} \,+\, \frac{1}{4}\,\left(d\theta^{3}+\omega\right)^{2}\ ,
\end{equation}
where the round $S^{2}$ is parametrized by $(\theta^{1},\theta^{2})$, and $d\omega\,=\,\mathrm{vol}_{S^{2}}$. The above metric can be viewed as a (trivial) lens space bearing a unit NUT charge \cite{Cvetic:2000cj}.
Hence it becomes very natural to deform the range of the fiber coordinate $\theta^{3}$ by turning on a non-trivial NUT charge.
This procedure yields the brane system depicted in table~\ref{Table:BONUT}, which turns out preserve the same amount of supersymmetry as the one in table~\ref{Table:BO}.
\begin{table}[h!]
\renewcommand{\arraystretch}{1}
\begin{center}
\scalebox{1}[1]{
\begin{tabular}{c||c c c c c || c c||c c c}
branes & $t$ & $y^{1}$ & $y^{2}$ & $y^{3}$ & $y^{4}$ & $z$ & $\rho$ & $\theta^{1}$ & $\theta^{2}$ & $\theta^{3}$ \\
\hline 
D8 & $\times$ & $\times$ & $\times$ & $\times$ & $\times$ & $-$ & $\times$ & $\times$ & $\times$ & $\times$ \\
D4 & $\times$ & $\times$ & $\times$ & $\times$ & $\times$ & $-$ & $-$ & $-$ & $-$ & $-$ \\
KK5 & $\times$ & $\times$ & $\times$ & $\times$ & $\times$ & $\times$ & $-$ & $-$ & $-$ & ISO 
\end{tabular}
}
\end{center}
\caption{{\it The brane picture underlying 5d $\ma N=2$ SCFT's defined by the D4-D8-KK5 system. The system is $\text{BPS}/4$ and
in the $\textrm{AdS}_{6}\times S^4/\mathbb{Z}_{k}$ vacuum the AdS radial coordinate is represented by a combination of $\rho$ and $z$, while the 
$\mathbb{Z}_{k}$ orbifold is realized by the KK5 charge.}} \label{Table:BONUT}
\end{table}

The massive type IIA supergravity background describing a semilocalized D4-D8-KK5 system reads
\begin{equation}
   \begin{split}
    d s_{10}^2&=H_{\mathrm{D}8}^{-1/2}\,H_{\mathrm{D}4}^{-1/2}\,d s^2_{\mathbb{R}^{1,4}}+H_{\mathrm{D}8}^{1/2}\,H_{\mathrm{D}4}^{1/2}\,dz^2+H_{\mathrm{D}8}^{-1/2}\,H_{\mathrm{D}4}^{1/2}\,H_{\mathrm{KK}5}\,\left(d \rho^2+\rho^2\,d s^2_{S^2} \right)\,+\\
   &\phantom{=}\,+\,H_{\mathrm{D}8}^{-1/2}\,H_{\mathrm{D}4}^{1/2}\,H_{\mathrm{KK}5}^{-1}\,\left(d\theta^{3}+Q_{\textrm{KK}5}\,\omega\right)^{2} \ ,\\
   e^{\Phi}&=g_s\,H_{\mathrm{D}8}^{-5/4}\,H_{\mathrm{D}4}^{-1/4}\ ,\qquad C_{(5)}=\frac{1}{g_s\,H_{\mathrm{D}4}}\ ,
   \label{BOsolution}
   \end{split}
  \end{equation}
where $H_{\mathrm{D}8}=H_{\mathrm{D}8}(z)$, $H_{\mathrm{D}4}=H_{\mathrm{D}4}(z,\rho)$ and $H_{\mathrm{KK}5}=H_{\mathrm{KK}5}(\rho)$ are suitable functions given by
\begin{equation}
H_{\mathrm{D}4}(z, \rho)=1+\frac{Q_{{\scriptsize{\mrm{D}4}}}}{(\rho+\frac{g_s m}{9Q_{\mathrm{KK}5}}\,z^3)^{5/3}}\ ,\quad H_{\mathrm{D}8}(z)=g_s\,m\,z
\quad \text{and}\quad H_{\mathrm{KK}5}(\rho)=\frac{Q_{\mathrm{KK}5}}{\rho}\ .
\label{BOsolutionKK}
\end{equation}
If we now introduce 
\begin{equation}
 \rho=\frac{g_s m}{9}\,\zeta^{3}\,\cos^{2}\alpha \qquad \text{and}\qquad z=Q_{\mathrm{KK}5}^{1/3}\,\zeta\,\sin^{2/3}\alpha\ ,
\end{equation}
the metric \eqref{BOsolutionKK} takes the form
\begin{equation}
\ell^{2}\,ds_{10}^{2}\,=\, s^{-1/3}\,\left(ds_{\mathrm{AdS}_{6}}^{2}\,+\,\frac{4}{3^{5/3}}\,Q_{\mathrm{KK}5}\,ds^{2}_{S^{4}/\mathbb{Z}_{k}}\right)\ ,
\end{equation}
with $\ell^{2}=3^{5/3}\,(g_s m)^{1/3}Q_{\mathrm{KK}5}^{1/6}Q_{\mathrm{D}4}^{-1/2}$ and
\begin{equation}
 ds^{2}_{S^{4}/\mathbb{Z}_{k}}=d\alpha^2+\frac{c^2}{4}\left(d s^2_{S^2} +\left(Q_{\mathrm{KK}5}^{-1}\,d\theta^{3}+\omega\right)^{2}\right)\,,
\end{equation}
where $s=\sin\alpha$ and $c=\cos\alpha$.

\section{The Supergravity Setup}

The bosonic isometries of the $\mrm{AdS}_6\times S^4$ vacuum \eqref{BOvacuum} introduced in section \ref{D4D8system} are naturally embedded into the $F(4)$ superalgebra and this hints a strong link with minimal\footnote{By ``minimal'' we mean the truncation to the pure supergravity multiplet of the theory.} $\ma N=(1,1)$ gauged supergravity in $d=6$. This theory is also known as $F(4)$ or Romans supergravity and it was firstly studied in \cite{Romans:1985tw}. In this section we will introduce the main properties of this supergravity theory, we will present the unique supersymmetric $\mathrm{AdS}_6$ vacuum admitted by the scalar potential and we will revisit some domain wall solutions as simplest examples of backgrounds involving non-trivial field profiles.

Subsequently we will present the consistent truncation of massive IIA supergravity around the $\mrm{AdS}_6\times S^4$ \cite{Cvetic:1999un}. This will turn out to reproduce exactly the equations of motion of $F(4)$ gauged supergravity. For this reason this 6d supergravity will constitute a powerful tool to capture the low-energy physics of those brane systems in massive IIA that are related to the D4-D8 constructions presented in section \ref{D4D8system}.

\subsection{Minimal $\ma N=(1,1)$ Gauged Supergravity in $d=6$}

Half-maximal $\ma N=(1,1)$ supergravities in $d=6$ \cite{Romans:1985tw,Andrianopoli:2001rs} admit the coupling of the supergravity multiplet to an arbitrary number $n$ of matter multiplets. Each of these includes four real scalar fields and the entire set of moduli parametrizes the $(4n+1)$-dimensional coset
\begin{equation}
 \mathbb R^+ \times \frac{\mathrm{SO}(4,n)}{\mathrm{SO}(4)\times \mathrm{SO}(n)}\ .
\end{equation}
In this paper we consider the minimal realization of $\ma N=(1,1)$ supergravity in $d=6$, then retaining in our analysis only the pure supergravity multiplet. 
We refer to appendix~\ref{app:halfmax} for the details of the truncation yielding the theory in its minimal incarnation as originally introduced in \cite{Romans:1985tw}.
In this case the global isometry group breaks down to \cite{Andrianopoli:2001rs,Karndumri:2016ruc,Karndumri:2012vh}
\begin{equation}
 G_0=\mathbb R^+ \times \mathrm{SO}(4)\,.
 \label{global}
\end{equation}
The R-symmetry group is the diagonal $\mathrm{SU}(2)_R \subset \mathrm{SO}(4)\simeq \mathrm{SU}(2)\times \mathrm{SU}(2)$ corresponding to $16$ preserved supercharges, which are in turn organized in their irreducible chiral components. 
The fermionic field content of the supergravity multiplet is given by two gravitini and two gaugini. Both the gravitini and the gaugini can be packed into pairs of Weyl spinors with opposite chiralities. Moreover, in $d=1+5$ spacetimes it is possible to introduce symplectic-Majorana-Weyl spinors\footnote{For more details on Clifford algebras for $d=1+5$ spacetime dimensions see Appendix~\ref{app:SM_spinors}.} (SMW). This formulation turns out to be very convenient in that it arranges the fermionic degrees of freedom of the theory into $\mathrm{SU}(2)_R$ doublets, respectively denoted by $\psi_{\mu}^a$ and $\chi^a$ with $a=1,2$. Note that such objects must also respect the pseudo-reality condtion \eqref{SM_cond} in order for them to describe the correct number of propagating degrees of freedom.

The bosonic content of the supergravity multiplet consists of the graviton $e_\mu^m$ with $m=0,\dots,5$, a real scalar $X$, a 2-form gauge potential $\ma{B}_{(2)}$, a non-Abelian $\mathrm{SU}(2)$ valued vector $A^i$ and an Abelian vector $A^0$.

The consistent deformations of the minimal theory are determined by the gauging of the R-symmetry $\mathrm{SU}(2)_R \subset \mathrm{SO}(4)$, through the vectors $A^i$, and by a St\"uckelberg coupling giving mass to the 2-form $\ma{B}_{(2)}$. The first deformation is described by a coupling constant $g$ and the second by a mass parameter $m$.

The bosonic Lagrangian has the form \cite{Romans:1985tw,Cvetic:1999un,Nunez:2001pt}
\begin{equation}
\begin{split}
\label{6dlagrangian}
 \ma L&=  R\,\star_{\,6}  1-4 \, X^{-2}\,\star_{\,6}\,dX\,\wedge \, dX-\frac12\,X^4\,\star_{\,6}\,\ma F_{(3)}\wedge \ma F_{(3)}-V(X)\\
 &- \frac12\, X^{-2} \,\left( \,\star_{\,6}\,\ma{F}_{(2)}^{i}\wedge \ma{F}_{(2)}^i+\star_{\,6}\,\ma{H}_{(2)}\wedge \ma{H}_{(2)}\,\right)-\frac12 \, \ma{B}_{(2)}\,\wedge \,\ma F_{(2)}^0\,\wedge \ma F_{(2)}^0\\
 &-\frac{1}{\sqrt 2}\,m \, \ma{B}_{(2)}\,\wedge \,\ma{B}_{(2)}\,\wedge \ma F_{(2)}^0-\frac13\,m^2 \, \ma{B}_{(2)}\,\wedge \,\ma{B}_{(2)}\,\wedge \ma{B}_{(2)}-\frac12 \, \ma{B}_{(2)}\,\wedge\, \ma{F}_{(2)}^{i}\wedge \ma{F}_{(2)}^i\,,
 \end{split}
\end{equation}
where the field strengths are defined as
\begin{equation}
 \begin{split}
  \mathcal{F}_{(3)} \ &= \ d\ma{B}_{(2)} \ ,\\
  \mathcal{F}_{(2)}^0 \ &= \ dA^0 \ ,\\
  \mathcal{H}_{(2)} \ &= \ dA^0+\sqrt{2}\,m\,\ma{B}_{(2)} \ ,\\
  \mathcal{F}_{(2)}^{i} \ &= \  d A^{i}  + \frac{g}{2} \, \epsilon^{ijk}\,A^{j}\wedge A^{k} \ .\\
 \end{split}
\end{equation}
The scalar potential $V(X)$ induced by the gauging is given by
\begin{equation}
 V(X)=m^2\,X^{-6}-4\sqrt{2}\,gm\,X^{-2}-2\,g^2\,X^2\ ,
 \label{scalarpotential}
\end{equation}
and it can be expressed in terms of a real function $f(X)$, the BPS superpotential, as it follows
\begin{equation}
 V(X)  = 16\,X^{2}\,\left(D_{X}f\right)^{2} -80\,f(X)^{2}\ ,
\end{equation}
where $f(X)$ is given by
\begin{equation}
 f(X)=\frac{1}{8}\,\left(m\,X^{-3}+\sqrt2 \,g\, X \right)\ .
 \label{superpotential}
\end{equation}
The SUSY variations of the fermions are expressed in terms of a 6d Killing spinor $\zeta^a$ in the following way \cite{Romans:1985tw,Nunez:2001pt}
\begin{equation}
 \begin{split}
 \label{SUSYvariation6d}
  \delta_\zeta \psi_{\mu}^a\,=\,&\,\nabla_\mu\,\zeta^a+4g\,(A_\mu)^{a}_{\,\,\,b}\,\zeta^b+\frac{X^2}{48}\,\Gamma_*\Gamma^{mnp}\,\ma F_{(3)\,mnp}\,\Gamma_\mu\zeta^a\\
  &+i\,\frac{X^{-1}}{16\sqrt2}\,\left(\Gamma_\mu^{\,\,\,mn}-6\,e_\mu^m\,\Gamma^n\right)\,(\ma{\hat{H}}_{mn})^{a}_{\,\,\,b}\,\zeta^b-if(X)\,\Gamma_\mu\Gamma_*\zeta^a\ ,\\
  \delta_\zeta \chi^{a}\,=\,&\,X^{-1}\Gamma^m\partial_m X\,\zeta^a+\frac{X^2}{24}\,\Gamma_*\Gamma^{mnp}\,\ma F_{(3)\,mnp}\,\zeta^a\\
  &-i\,\frac{X^{-1}}{8\sqrt2}\,\Gamma^{mn}(\ma{\hat{H}}_{mn})^{a}_{\,\,\,b}\,\zeta^b+2i\,XD_X\,f(X)\,\Gamma_*\,\zeta^a\ ,
 \end{split}
\end{equation}
with $\nabla_\mu\,\zeta^a=\partial_\mu\zeta^a+\frac14\,\omega_\mu^{\,\,\,mn}\,\Gamma_{mn}\,\zeta^a$ and $(\ma{\hat{H}}_{mn})^{a}_{\,\,\,b}$ defined as
\begin{equation}
 (\ma{\hat{H}}_{\mu\nu})^{a}_{\,\,\,b}=\ma H_{(2)\,\mu\nu}\,\delta^{a}_{\,\,\,b}-4\,\Gamma_*\,(\ma F_{(2)\,\mu\nu})^{a}_{\,\,\,b}\ ,
\end{equation}
where we introduced the notation $A^{a}_{\,\,\,b}=\frac12\,A^i(\sigma^i)^{a}_{\,\,\,b}$ with $\sigma^i$ Pauli matrices given in \eqref{Pauli}.
Varying \eqref{6dlagrangian} with respect to all the bosonic fields we obtain the equations of motion
\begin{equation}
 \begin{split}
 &R_{\mu\nu}-4\,X^{-2}\,\partial_\mu X\,\partial_\nu\,X-\frac14\,V(X)\,g_{\mu\nu}-\frac14\,X^4\,\left(\ma F_{(3)\,\mu}^{\quad \,\,\, \alpha\beta}\ma F_{(3)\,\nu\alpha\beta}-\frac16\,\ma F_{(3)}^2\,g_{\mu\nu}\right)\\\vspace{2mm}
  &-\frac12\,X^{-2}\,\left(\ma H_{(2)\,\mu}^{\quad\,\,\,\alpha\,}\ma H_{(2)\,\nu\alpha}-\frac18\,\ma H_{(2)}^2g_{\mu\nu}\right)-\frac12\,X^{-2}\left(\ma F_{(2)\,\mu}^{i\quad\alpha}\,\ma F^i_{(2)\,\nu\alpha}-\frac18\ma F_{(2)}^{i\,\,\,2}\,g_{\mu\nu}\right)=0\,,\\\vspace{2mm}
  &d\left(X^4\,\star_{\,6}\,\ma F_{(3)}\right)=-\frac12\,\ma H_{(2)}\,\wedge\,\ma H_{(2)}-\frac12\,\ma F^i_{(2)}\,\wedge\,\ma F^i_{(2)}-\sqrt{2}\,m \,X^{-2}\,\star_{\,6}\,\ma H_{(2)}\,,\\\vspace{2mm}
  &d\left(X^{-2}\,\star_{\,6}\,\ma H_{(2)}\right)=-\ma H_{(2)}\,\wedge\,\ma F_{(3)}\,,\\\vspace{2mm}
  & D\left( X^{-2}\,\star_{\,6}\,\ma F_{(2)}^i\right)=-\ma F^i_{(2)}\,\wedge\,\ma F_{(3)}\,,\\\vspace{3mm}
 &d\left( X^{-1}\star_{\,6}dX\right)+\,\frac18\,X^{-2}\,\left(\star_{\,6}\,\ma H_{(2)}\,\wedge\,\ma H_{(2)}+\star_{\,6}\,\ma F^i_{(2)}\,\wedge\,\ma F^i_{(2)}\right)\\
 &-\frac14\,X^4\,\star_{\,6}\ma F_{(3)}\,\wedge\,\ma F_{(3)}-\frac18\,X\,D_X\,V(X)\star_{\,6}1=0\,,
 \end{split}
 \label{eom}
\end{equation}
where $D$ is the gauge covariant derivative defined as $D \,\omega^i=d\omega^i+g\,\epsilon_{ijk}\,A_{}^j\,\wedge\,\omega^k$ with $\omega^i$ any $\mathrm{SU}(2)$ covariant quantity.

\subsection{$\mrm{AdS}_6$ Vacuum and Domain Walls}

The scalar potential \eqref{scalarpotential} admits a critical point giving rise to an $\mathrm{AdS}_6$ vacuum preserving 16 real supercharges. This vacuum is realized by the following value of $X$
\begin{equation}
 X=\frac{3^{1/4}\,m^{1/4}}{2^{1/8}\,g^{1/4}}\ ,
 \label{AdS6}
\end{equation}
and by setting all the gauge potentials to zero. 
The simplest excited background in 6d $\ma N=(1,1)$ gauged supergravity is a field configuration involving only the scalar $X$. Such a system is described by a domain wall flow of the type
\begin{equation}
 \begin{split}
 & ds_{6}^{2}  =  e^{2V(r)} \,dr^{2}+ e^{2U(r)}\,ds_{\mathbb{R}^{1,4}}^{2} \ , \\
& X =  X(r) \,,
\label{DWansatz}
 \end{split}
\end{equation}
where $ds_{\mathbb{R}^{1,4}}^{2}$ is the metric of the 5d Minkowski spacetime. In order to derive the explicit radial dependence of the warp factors and of the scalars, we can set to zero the SUSY variations of fermions \eqref{SUSYvariation6d} and choose as Killing spinor a Dirac spinor\footnote{The fermionic parameter $\zeta^a$ appears inside the SUSY variations \eqref{SUSYvariation6d} as a SMW spinor since vector fields have a natural $\mathrm{SU}(2)$ action on spinor doublets. As we explained in appendix~\ref{app:SM_spinors}, the pseudo-reality condition \eqref{SM_cond} guarantees that the number of independent components of a SM (SMW) doublet are the same as those of a Dirac (Weyl) spinor. This means that, whenever vectors are vanishing, it will be more suitable to reorganize them into Dirac or Weyl spinors. } $\zeta$ of the form 
\begin{equation}
\zeta(r) = Y(r)\,\zeta_{0}\ ,
\label{KspinorDW}
\end{equation}
where $\zeta_{0}$ is a constant Dirac spinor satisying the projection condition
\begin{equation}
 -i\,\Gamma^3\,\Gamma_*\,\zeta_{0}=\zeta_{0}\ .
 \label{projDW}
\end{equation}
Imposing the background \eqref{DWansatz} with the Killing spinor \eqref{KspinorDW}, the SUSY variations \eqref{SUSYvariation6d} reduce to a set of flow equations given by
\begin{equation}
 U' = -2 \, e^{V} \, f(X) \ , \qquad Y' \, = -Y\, e^{V} \,f(X) \ , \qquad X'  = 2\, e^{V} \,X^{2} \,D_{X}f \ .
 \label{floweqDW}
\end{equation}
The warp factor $V$ is pure gauge and it can be defined as
\begin{equation}
 e^{V}=\frac{ X^{-2}}{2\, D_{X}f}\ ,
\end{equation}
so that the flow equations \eqref{floweqDW} can be easily intergrated to give
\begin{equation}
 e^{2U}= \left(\frac{r}{\sqrt{2}\, g \,r^4-3 \,m}\right)^{2/3}\  , \qquad e^{2V}= \left(\frac{4\,r^2}{\sqrt{2}\, g \,r^4-3 \,m}\right)^{2}\,,\qquad X=r\ ,
 \label{DWsolution}
\end{equation}
with a radial dependence of the Killing spinor specified by $Y=e^{U/2}$.

\subsection{The Massive IIA Origin of $F(4)$ Supergravity}
\label{reduction}

In this subsection we present the consistent truncation of massive IIA supergravity around the $\mrm{AdS}_6\times S^4$ vacuum introduced in section \ref{D4D8system}. The 6d vacuum \eqref{AdS6} will then gain a natural interpretation in massive IIA string theory as the near-horizon of a D4-D8 system and $F(4)$ gauged supergravity will turn out to be the effective theory capturing the physics associated to the background's excitations around this vacuum.
The stringy interpretation of \eqref{AdS6} is realized thanks to the reduction Ansatz constructed in \cite{Cvetic:1999un}, in which a consistent truncation to the theory \eqref{6dlagrangian} is constructed. In particular, after fixing the 6d gauge parameter as
\begin{equation}
m \ = \ \frac{\sqrt{2}\,g}{3}\ ,
\label{m&guplift}
\end{equation}
 the 6d equations of motion \eqref{eom} can be obtained from the following truncation Ansatz of the 10d background\footnote{For our later convenience, we formulate the Ansatz in the string frame, while in \cite{Cvetic:1999un} it is given in the Einstein frame. See appendix \ref{app:massiveIIA}.} \cite{Cvetic:1999un}
\begin{equation}
 \begin{split}
  ds^2_{10}=s^{-1/3}\,X^{-1/2}\,\Delta^{1/2}\,\left[ds^2_6+2g^{-2}\,X^{2}\,ds_{\tilde{S}^4}^2\right]\ ,
  \label{truncationansatz}
 \end{split}
\end{equation}
where $\Delta=Xc^2+X^{-3}s^2$ and $ds_{\tilde{S}^4}^2$ is the metric of a squashed 4-sphere $\tilde{S}^4$ describing a fibration of a 3-sphere over a circle
\begin{equation}
ds_{\tilde{S}^4}^2=d\xi^2+\frac14\,\Delta^{-1}\,X^{-3}\,c^{2}\,\sum_{i=1}^3\,\left(\theta^i-gA^i  \right)^2\ ,
 \label{4-sphere}
\end{equation}
with $c=\cos\xi$ and $s=\sin \xi$. By observing the internal structure of \eqref{4-sphere}, one may immediately conclude that also the internal 3-sphere is deformed and, in particular, it identifies an $\mathrm{SU}(2)$ bundle for which the 6d vectors $A^i$ are the connections and $\theta^i$ the left-invariant 1-forms\footnote{They satisfy the identity $d\theta^i=-\frac12\,\varepsilon_{ijk}\,d\theta^j\,\wedge\,d\theta^k$.}.
The rest of the 10d fields are given by \cite{Cvetic:1999un}
\begin{equation}
\label{10dfluxes}
 \begin{split}
  F_{(4)}&=-\frac{\sqrt 2}{6}\,g^{-3}\,s^{1/3}\,c^3\,\Delta^{-2}\,U\,d\xi\,\wedge\,\epsilon_{(3)}-\sqrt{2}\,g^{-3}\,s^{4/3}\,c^4\,\Delta^{-2}\,X^{-3}\,dX\,\wedge\,\epsilon_{(3)}\\
  &-\sqrt2 \,g^{-1}\,s^{1/3}\,c\,X^4\,\star_{\,6}\ma F_{(3)}\,\wedge\,d\xi-\frac{1}{\sqrt 2}\,s^{4/3}\,X^{-2}\,\star_{\,6}\ma H_{(2)}\\
  &+\frac{g^{-2}}{\sqrt 2}\,s^{1/3}\,c\,\ma F_{(2)}^i\,h^i\,\wedge\,d\xi-\frac{g^{-2}}{4\sqrt2}\,s^{4/3}\,c^2\,\Delta^{-1}\,X^{-3}\,\epsilon_{ijk}\,\ma F_{(2)}^i\,\wedge h^j\wedge\,h^k\ ,\\
  F_{(2)}&=\frac{s^{2/3}}{\sqrt 2}\,\ma H_{(2)}\ ,\qquad H_{(3)}=s^{2/3}\,\ma F_{(3)}+g^{-1}\,s^{-1/3}\,c\,\ma H_{(2)}\,\wedge\,d\xi\ ,\\
  e^{\Phi}&=s^{-5/6}\,\Delta^{1/4}\,X^{-5/4}\ ,\qquad F_{(0)}=m\ .
 \end{split}
\end{equation}
where $U=X^{-6}\,s^2-3X^2\,c^2+4\,X^{-2}\,c^2-6\,X^{-2}$ and $\epsilon_{(3)}=h^1\,\wedge\,h^2\,\wedge\,h^3$ with $h^i=\theta^i-gA^i$.
Expressing \eqref{AdS6} in terms of \eqref{m&guplift}, one obtains the $\mrm{AdS}_6\times S^4$ vacuum \eqref{BOvacuum}. In particular, for $X=1$ and vanishing gauge potentials, the manifold \eqref{4-sphere} becomes a round 4-sphere\footnote{As pointed out in \cite{Brandhuber:1999np} and in the discussion above on \eqref{BOvacuum}, this is only the upper hemisphere of a 4-sphere with a bounday appearing for $\xi\rightarrow 0$.}. From \eqref{10dfluxes} it follows that $F_{(4)}$ is the only non-zero flux, in addition to the Romans' mass, supporting the $\mrm{AdS}_6\times S^4$ vacuum. Together with the dilaton, it has the following form
\begin{equation}
 F_{(4)}=\frac{5\sqrt{ 2}}{6}\,g^{-3}\,s^{1/3}\,c^3\,d\xi\,\wedge\,\epsilon_{(3)}\ ,\qquad e^{\Phi}=s^{-5/6}\ ,
\end{equation}
which are exactly the flux and dilaton configurations corresponding to the near-horizon of the semilocalized D4-D8 system introduced in section \ref{D4D8system} \cite{Brandhuber:1999np, Cvetic:1999un}.

In terms of an embedding tensor/fluxes dictionary, the massive type IIA origin of the minimal theory is summarized in table~\ref{Table:ET/fluxes}.
\begin{table}[h!]
\renewcommand{\arraystretch}{1}
\begin{center}
\scalebox{1}[1]{
\begin{tabular}{| c | c | c | c |}
\hline
Fluxes & $\Theta$ & Minimal & $\mathbb{R}^{+}_{X}$ weights \\
\hline
$F_{0ijk}$ & $\zeta_{0}\,\epsilon_{ijk}$ & $m$ & $+3$ \\[1mm]
\hline 
$F_{(0)}$ & $\frac{1}{3!}\epsilon^{ijk}\,f_{ijk}$ & $g$ & $+1$ \\[1mm]
\hline
$\omega_{ij}^{\phantom{kl}k}$ & $f_{ijk}$ & $g$ & $+1$ \\[1mm]
\hline
\end{tabular}
}
\end{center}
\caption{{\it The embedding tensor/fluxes dictionary specifying the massive IIA origin of Romans' theory in 6d. The $\Theta$ notation refers to the theory coupled to four vector multiplets in appendix~\protect\ref{app:halfmax}. $\omega_{ij}{}^{k}$ refers to the spin connection of $S^{3}$.}} \label{Table:ET/fluxes}
\end{table}
Note that this massive IIA realization of Romans' theory supports spacetime-filling KK monopoles. As already mentioned in appendix~\ref{app:halfmax}, the presence of such a tadpole is inferred by a violation of the extra constraints in \eqref{extraQC}.
The fact that the source is of a KK5-brane type is due to the fact that its WZ action is constructed through the coupling to a \emph{mixed} symmetry potential of $(7,1)$ type.
The corresponding tadpole will then be a $(3,1)$-form. Such an object can be constructed in our case as $\theta^{ae}\,F_{bcde}$, where $a$, $b$, $c$ and $d$ are $\mathrm{SO}(4)$ indices and 
$\theta^{ij}$ is constructed from the above $\omega_{ij}^{\phantom{kl}k}$ by contracting it with $\epsilon^{ijk}$.
Such KK5 branes as spacetime-filling sources exactly correspond to the objects appearing in the brane system introduced in table~\ref{Table:BONUT}.

In the following section we are going to present new classes of solutions to 6d $F(4)$ supergravity involving non-trivial profiles for the two-form field.
Thanks to the uplift formulae revisited in this section, these will gain a natural massive type IIA origin that will allow us to speculate on their possible holographic interpretation.

\section{BPS Flows with the 2-form Gauge Potential}

\label{flows}

In this section we derive a new class of supersymmetric solutions for the theory \eqref{6dlagrangian} by solving the BPS equations associated to the SUSY variations \eqref{SUSYvariation6d}. These flows are characterized by a non-trivial profile for the 2-form gauge potential $\ma{B}_{(2)}$ and some of them enjoy a UV regime reproducing locally the $\mrm{AdS}_6$ vacuum \eqref{AdS6}. The spacetime backgrounds defining these solutions may be divided into two classes: one featured by a three-dimensional Minkowski $\mathbb R^{1,2}$ slicing and the other by a $\mrm{AdS}_3$ foliation.

We will firstly formulate the general Ansatz on the bosonic fields and on the Killing spinor giving rise to the first-order flow associated to this class of backgrounds. Then we will explicitly solve the first-order equations obtaining a class of novel solutions preserving 8 real supercharges.

\subsection{The General Ansatz}

The 6d metrics considered are of the general form
\begin{equation}
 ds_6^2=e^{2U(r)}\,ds^2_{M_3}+e^{2V(r)}\,dr^2+e^{2W(r)}\,ds^2_{\Sigma_2}\ ,
 \label{general6dmetric}
\end{equation}
where the ``worldvolume'' part $M_3$ is given by the 3-dimensional Minkowski spacetime $\mathbb{R}^{1,2}$ or by $\mrm{AdS}_3$, and the ``transverse'' space $\Sigma_2$ can be either $\mathbb{R}^2$ or $S^2$.
As in the case of the domain wall solution \eqref{DWsolution}, we introduce the non-dynamical warp factor $V$ that will turn out to be crucial to analytically solve the flow equations.

For simplicity we will consider vanishing vectors, \emph{i.e.} $A^i=0$ and $A^0=0$ and, as far as the 2-form gauge potential $\ma{B}_{(2)}$ is concerned, it will be considered wrapping the manifold $\Sigma_2$ as it follows
\begin{equation}
 \ma{B}_{(2)}=b(r)\,\text{vol}_{\Sigma_2}\ .
 \label{general2form}
\end{equation}
We furthermore also assume a purely radial dependence for the scalar
\begin{equation}
 X=X(r)\ .
\end{equation}
Since we are looking for SUSY backgrounds, we need to specify a suitable Killing spinor realizing a set of non-trivial first-order equations corresponding to the spacetime background given in \eqref{general6dmetric} and \eqref{general2form}. As in the case of the domain wall \eqref{KspinorDW}, the action of the SUSY variations on the $\mathrm{SU}(2)_R$ indices of the Killing spinor $\zeta^a$ is trivial, so it is more natural to reorganize the components of a Killing spinor into a $(1+5)$-dimensional Dirac spinor $\zeta$.
Following the splitting of the Clifford algebra given in \eqref{gammamatrices}, the Killing spinors considered are of the form
\begin{equation}
\begin{split}
 \zeta(r)=&\ \zeta^++i\,B\,\Gamma_*\,\zeta^-\ ,\\
 \zeta^\pm =& \,Y(r)\,\eta_{M_3}\,\otimes \,\left(\cos\theta(r)\,\chi^\pm_{\Sigma_2}\otimes \varepsilon_0+i\,\sin\theta(r)\,\gamma_*\,\chi^\pm_{\Sigma_2}\,\otimes\sigma^3 \varepsilon_0  \right)\ ,
 \label{generalKilling}
 \end{split}
\end{equation}
where the explicit representations of the chiral operator $\Gamma_*$ is defined in \eqref{abcrep} and the complex-conjugation matrix $B$ in \eqref{gammastarrep} in terms of the Dirac matrices \eqref{gammai} on $\Sigma_2$.
The spinor $\eta_{{\scriptsize\mrm{AdS}_3}}$ on $M_3=\mrm{AdS}_3$ is a Majorana Killing spinor enjoying 2 real independent components and satisfying the following Killing equation
\begin{equation}
 \nabla_{\scriptsize{x^\alpha}}\, \eta_{M_3}=\frac{L}{2}\,\rho_{\scriptsize{x^\alpha}}\,\eta_{M_3}\ ,
 \label{killingM3}
\end{equation}
where $\rho_{\scriptsize{x^\alpha}}$ are the Dirac matrices introduced in \eqref{rho3d} and $L^{-1}$ the radius of $\mrm{AdS}_3$. The flat case $M_3=\mathbb{R}^{1,2}$ is recovered by taking a solution of \eqref{killingM3} with $L=0$.

Let us now consider the Euclidean spinor $\chi_{S^2}$ on $\Sigma_2=S^2$ with radius $R^{-1}$. This is a complex spinor carrying 4 real independent degrees of freedom that can split into 2+2 components $\chi^\pm_{S^2}$ solving the following Killing conditions on $S^2$,
\begin{equation}
\begin{split}
& \nabla_{\scriptsize{\theta^i}}\, \chi^+_{\Sigma_2}=\frac{R}{2}\,\gamma_*\,\gamma_{\scriptsize{\theta^i}}\,\chi^-_{\Sigma_2}\ ,\\
& \nabla_{\scriptsize{\theta^i}}\, \chi^-_{\Sigma_2}=\frac{R}{2}\,\gamma_*\,\gamma_{\scriptsize{\theta^i}}\,\chi^+_{\Sigma_2}\ .
\label{killingS2}
 \end{split}
\end{equation}
In the $R=0$ limit we obtain the Killing spinor equations for the flat case $\Sigma_2=\mathbb{R}^2$ in which $\chi_{\mathbb{R}^2}^+=\chi_{\mathbb{R}^2}^-\equiv \chi_{\mathbb{R}^2}$.

Finally $\varepsilon_0$ is a 2-dimensional real constant spinor encoding the two different chiral parts of $\zeta$ as
\begin{equation}
 \Gamma_*\,\zeta=\pm\,\zeta \qquad \Longleftrightarrow \qquad \sigma^3\,\varepsilon_0=\pm\,\varepsilon_0\ ,
\end{equation}
where we used the identity \eqref{gammastarrep}. Summarizing, we have that our $\zeta$ depends on 16 real independent components in total. As we shall see later, these will be reduced by half by an algebraic projection condition associated with the particular background considered.

\subsection{Background with $M_3=\mathbb{R}^{1,2}$ and $\Sigma_2=\mathbb{R}^{2}$}
\label{sec:MkW3R2}

Let's start with the simplest configuration in which the metric \eqref{general6dmetric} is featured by $M_3=\mathbb{R}^{1,2}$ and $\Sigma_2=\mathbb{R}^{2}$. The 6d background takes the following form
\begin{equation}
 \begin{split}
  &ds_6^2=e^{2U(r)}\,ds^2_{\mathbb{R}^{1,2}}+e^{2V(r)}\,dr^2+e^{2W(r)}\,ds^2_{\mathbb{R}^{2}}\ ,\\
  &\ma{B}_{(2)}=b(r)\,\text{vol}_{\mathbb{R}^{2}}\ ,\\
  & X=X(r)\,.
  \label{MkW3R2ansatz}
 \end{split}
\end{equation}
The Killing spinor realizing the background \eqref{MkW3R2ansatz} is included into the general expression given in \eqref{generalKilling}. In the case in which both $M_3$ and $\Sigma_2$ are flat, the spinors $\eta_{\mathbb{R}^{1,2}}$, $\chi_{\mathbb{R}^2}^\pm$ respectively satisfy the Killing spinor equations \eqref{killingM3} and \eqref{killingS2} in the limits where both $L=0$ and $R=0$. This implies that the Killing spinor of the background \eqref{MkW3R2ansatz} may be written as
\begin{equation}
 \zeta=Y(r)\,\eta_{\,\mathbb{R}^{1,2}}\otimes \left(\cos\theta(r)\,\chi_{\mathbb{R}^2}\otimes \varepsilon_0+i\,\sin\theta(r)\,\gamma_*\,\chi_{\mathbb{R}^2}\otimes\sigma^3 \,\varepsilon_0  \right)\ .
 \label{killingMkW3R2}
\end{equation}
The projection condition \eqref{projDW} expressed in terms of \eqref{gammamatrices} takes the form
\begin{equation}
 (\gamma_*\otimes \sigma^1)\,(\chi_{\mathbb{R}^2}\otimes \varepsilon_0)=\chi_{\mathbb{R}^2}\otimes \varepsilon_0\ ,
 \label{projMkW3R2}
\end{equation}
where we omitted the spinor's $\mathbb{R}^{1,2}$ part since the action of \eqref{projDW} on $\eta_{\mathbb{R}^{1,2}}$ is given by the identity. We can recast \eqref{killingMkW3R2} in the more compact form given by
\begin{equation}
 \zeta=Y(r)\left(\cos\theta(r)\,\mathbb{I}_8-\sin \theta(r)\,\Gamma^4\,\Gamma^5\,\Gamma_*  \right)\zeta_0\ ,
\end{equation}
where $\zeta_0$ is a constant Dirac spinor satisfying the condition $-i\,\Gamma^3\,\Gamma_*\,\zeta_{0}=\zeta_{0}\,$.

Evaluating the SUSY variations \eqref{SUSYvariation6d} onto the background \eqref{MkW3R2ansatz} and the Killing spinor \eqref{killingMkW3R2} satisfying \eqref{projMkW3R2}, we obtain the following set of first-order equations
\begin{equation}
 \begin{split}
   U^\prime=&\ -2\,e^{V}\,\frac{\cos(4\theta)}{\cos(2\theta)}\,f\ ,\\
  W^\prime=&\ 2\, e^{V} \,\frac{\cos (4 \theta )-2}{ \cos (2 \theta )}\,f\ ,\\
  b^\prime=&\ \frac{16}{X^2} \,e^{V+2 W}\, \sin (2 \theta )\,f\ ,\\
  X^\prime=&\ \frac{2\, e^{V}\, X}{\cos(2\theta)}\, \left(\cos(2\theta)\,X D_X \,f+2 \,(2\cos(2\theta)-1)\,\sin^2(2\,\theta)\,f\right)\ ,\\
  Y^\prime=&\, -Y\,e^{V}\,\frac{\cos(4\theta)}{\cos(2\theta)}\,f\ ,\\
    \theta^\prime=&\ 4 \,e^{V}  \,\sin (2 \theta )\,f\ .
\label{floweqMkW3R2}
  \end{split}
\end{equation}
For consistency the above equations have to be supplemented by the two constraints
\begin{equation}
 \begin{split}
  &b\overset{!}{=}\frac{8}{m}\,e^{2W}\,X\,\tan(2\theta)\,f\ ,\\
  &X\,D_X\,f+3\,f\overset{!}{=}0\ .
  \label{constraintsMkW3R2}
 \end{split}
\end{equation}
The second relation of \eqref{constraintsMkW3R2} implies that the flow \eqref{floweqMkW3R2} must be driven by the run-away superpotential given by
\begin{equation}
f=\frac{m}{8}\,X^{-3}\ .
\label{superpotentialMkW3R2}
\end{equation}
If \eqref{superpotentialMkW3R2} holds, than the expression of $b$ in \eqref{constraintsMkW3R2} is automatically compatible with \eqref{floweqMkW3R2}.
In order to intergrate the equations \eqref{floweqMkW3R2} we make the following gauge choice
\begin{equation}
 e^{V}=(4\,f)^{-1}\ .
\end{equation}
Starting from the equation for $\theta^\prime$ we can solve the whole system obtaining
\begin{equation}
 \begin{split}
  e^{2U}=&\ \sinh(4 r)^{1/4}  \coth(2 r)^{3/4}\ ,\\
  e^{2W}=&\ \sinh (4 r)^{1/4} \tanh (2 r)^{5/4}\ ,\\
  e^{2V}=&\ \frac{4}{m^2}\, \coth(2 r)^{3/4}\, \sinh(4 r)^{9/4}\ ,\\
   b=&\,-\frac{1}{\sqrt2}\cosh(2r)^{-2}\ ,\\
   X=&\ \sinh(4 r) ^{3/8}\coth (2 r)^{1/8}\ ,\\
      Y=&\ \sinh(4 r)^{1/16}  \coth(2 r) ^{3/16}\ ,\\
       \theta=&\ \arctan\left(e^{2 r}\right)\ .
   \label{flowMkW3R2}
 \end{split}
\end{equation}
The solution \eqref{flowMkW3R2} satisfies the equations of motion \eqref{eom} with a run-away scalar potential given by 
\begin{equation}
 V(X)=m^2\,X^{-6}\ .
 \label{run-away}
\end{equation}
The potential \eqref{run-away} does not admit critical points so \eqref{flowMkW3R2} cannot be asymptotically $\mathrm{AdS}_6$ for $r \rightarrow +\infty$, while in the IR regime $r \rightarrow 0$ the background becomes singular.

\subsection{Background with $M_3=\mathrm{AdS}_3$ and $\Sigma_2=\mathbb{R}^{2}$}
\label{sec:AdS3R2}

Let's now consider a curved worldvolume part $M_3=\mathrm{AdS}_3$, the 6d spacetime background takes the following form
\begin{equation}
 \begin{split}
  ds_6^2=&\  e^{2U(r)}\,ds^2_{\mathrm{AdS}_3}+e^{2V(r)}\,dr^2+e^{2W(r)}\,ds^2_{\mathbb{R}^{2}}\ ,\\
  \ma{B}_{(2)}=& \ b(r)\,\text{vol}_{\mathbb{R}^{2}}\ ,\\
   X=& \ X(r)\ .
  \label{AdS3R2ansatz}
 \end{split}
\end{equation}
As opposed to the previous case, a Killing spinor for \eqref{AdS3R2ansatz} has to produce the new contributions to the SUSY variations coming from the non-zero curvature of $\mrm{AdS}_3$. Considering the general form \eqref{generalKilling}, these contributions are encoded in $\eta_{\mathrm{AdS}_3}$ satisfying \eqref{killingM3} with $L\neq 0$. In order to simplify the derivation of BPS equations one may notice that the first-order formulation of the theory defined by \eqref{SUSYvariation6d} is gauge-dependent, \emph{i.e.} it depends explicitly on the spin connections of the background. This means that we can look for a parametrization of $\mrm{AdS}_3$ producing contributions in the SUSY variations\footnote{Such a {\itshape parallelized basis} does not clearly exist for every manifold. For example, in the next section we will consider $\Sigma_2=S^2$ and we will be forced to include a dependence on the coordinates of the $S^2$ into the Killing spinor.} that do not depend on the internal coordinates of $\mrm{AdS}_3$. This would allow us to keep the same Killing spinor of the flat case \cite{Dibitetto:2017tve}. The parametrization of $\mrm{AdS}_3$ giving rise to constant components of the spin connections in the flat basis is the Hopf fibration,
\begin{equation}
 \D s_{\mathrm{AdS}_3}^{2}=\frac{1}{4L^2}\left[ (dx^1)^2+\cosh^2x^1 (dx^2)^2-\left(d t-\sinh x^1 d x^2\right)^2   \right] \  ,
\label{AdS3}
\end{equation}
where the corresponding non-symmetric dreibein has the following form
\begin{equation}
  \begin{split}
&e^{0}  =  \frac{1}{2L} \,\left(\D t -\sinh x^1 \D x^2\right)\  ,\\
&e^{1}  =  \frac{1}{2L} \, \left(\cos t \,\D x^{1}\,-\,\sin t \cosh x^1 \,\D x^{2}\right) \ ,\\
&e^{2}  = \frac{1}{2L}\left(\cos t \cosh x^1\D x^2 + \sin t \, \D x^1 \right) \ .
\end{split}
\label{AdS3vielbein}
\end{equation}
The dreibein \eqref{AdS3vielbein} defines a constant spin connection in the flat basis. As a consequence, in this non-symmetric parametrization of $\mrm{AdS}_3$, we can keep the same form of the Killing spinor given in \eqref{killingMkW3R2} with the projection condition \eqref{projMkW3R2}.

Evaluating the Ansatz \eqref{AdS3R2ansatz} into the SUSY variations \eqref{SUSYvariation6d} with the Killing spinor \eqref{killingMkW3R2} satisfying \eqref{projMkW3R2}, we obtain the following set of first-order equations
\begin{equation}
 \begin{split}
  U^\prime=&\ -\frac{1}{4}\,e^{V}\,\cos(2\theta)^{-1}\left((3+5\cos(4\theta))\,f+2 \sin{(2\theta)}^2\,X\,D_Xf   \right)\ ,\\
  W^\prime=&\ -\frac{1}{4}\,e^{V}\,\cos(2\theta)^{-1}\left((7+\cos(4\theta))\,f-6 \sin{(2\theta)}^2\,X\,D_Xf   \right)\ ,\\
  b^\prime=&\ -\frac{2}{X^2} \,e^{V+2 W}\, \sin (2 \theta )\,\left(f+3\,X\,D_Xf     \right)\ ,\\
  X^\prime=&\ \frac14\, e^{V}\,\cos(2\theta)^{-1}\, X\, \left((-1+\cos(4\theta))\,f+ (5+3\cos(4\theta))\,X\,D_Xf   \right)\ ,\\
    Y^\prime=&\ -\frac{Y}{8}\,e^{V}\,\cos(2\theta)^{-1}\left((3+5\cos(4\theta))\,f+2 \sin{(2\theta)}^2\,X\,D_Xf   \right)\ ,\\
      \theta^\prime=&\ e^{V}  \,\sin (2 \theta )\,\left(f-X\,D_Xf     \right)\ .
\label{floweqAdS3R2}
  \end{split}
\end{equation}
where one has to impose the two additional constraints
\begin{equation}
 \begin{split}
  b\overset{!}{=}&\,\frac{2}{m} \,e^{2 W}\, \tan (2 \theta )\,X\,\left(f-X\,D_Xf\right)\ ,\\
  L\overset{!}{=}&\,- e^{U}\, \sin (2 \theta )\,\left(3\,f+X\,D_Xf\right)\ .
  \label{constraintsAdS3R2}
 \end{split}
\end{equation}
The relations in \eqref{constraintsAdS3R2} are automatically satisfied if $f$ coincides with the superpotential of the theory \eqref{superpotential}.
If we perform the gauge choice
\begin{equation}
 e^V=\left(f-X\,D_Xf\right)^{-1}\ ,
\end{equation}
we can analytically intergrate the system in \eqref{floweqAdS3R2}, obtaining the following solution
\begin{equation}
 \begin{split}
  e^{2U}=&\ 2\,\sinh(4 r)\ ,\\
  e^{2W}=&\ 2\,\sinh (2 r)^{2} \tanh (2 r)\ ,\\
   e^{2V}=&\ \frac{2^{5/4}\,3^{3/2}}{m^{1/2}\,g^{3/2}}\, \tanh(2 r)^{-3}\ ,\\
   b=&\ -\frac{2^{5/4}\,g^{1/2}}{3^{1/2}\,m^{1/2}}\sinh(2r)\, \tanh (2 r)^2\ ,\\
   X=&\ \frac{3^{1/4}\,m^{1/4}}{2^{1/8}\,g^{1/4}} \tanh (2 r)^{-1/2}\ ,\\
     Y=&\ 2^{1/4}\,\sinh(4 r)^{1/4}\ ,\\
        \theta=&\ \arctan\left(e^{2 r}\right)\ .
   \label{flowAdS3R2}
 \end{split}
\end{equation}
The equations of motion \eqref{eom} are satified by the flow \eqref{flowAdS3R2} if the radius of $\mrm{AdS}_3$ takes the following form
\begin{equation}
 L=2^{3/8}\,3^{1/4}\,(g^{3}\,m)^{1/4}\ ,
\end{equation}
with $g>0$ and $m>0$. 
In the asymptotic limit $r \rightarrow + \infty$ the background \eqref{flowAdS3R2} defines locally the $\mrm{AdS}_6$ vacuum introduced in \eqref{AdS6}. As for the $r \rightarrow 0$ limit, the solution is singular.

\subsection{Background with $M_3=\mathbb{R}^{1,2}$ and $\Sigma_2=S^{2}$}
\label{sec:MkW3S2}

Let's consider the specular case of transverse space with non-zero curvature, \emph{i.e.} $\Sigma=S^2$. In this case the 6d background takes the following form
\begin{equation}
 \begin{split}
  ds_6^2=&\ e^{2U(r)}\,ds^2_{\mathbb{R}^{1,2}}+e^{2V(r)}\,dr^2+e^{2W(r)}\,ds^2_{S^2}\ ,\\
  \ma{B}_{(2)}=&\ b(r)\,\text{vol}_{S^2}\ ,\\
   X=&\ X(r)\ ,
  \label{MkW3S2ansatz}
  \end{split}
\end{equation}
and the corresponding Killing spinor is given by
\begin{equation}
\begin{split}
 \zeta(r)=&\ \zeta^++i\,B\,\Gamma_*\,\zeta^-\ ,\\
 \zeta^\pm=&\ Y(r)\,\eta_{\mathbb{R}^{1,2}}\,\otimes \,\left(\cos\theta(r)\,\chi^\pm_{S^2}\otimes \varepsilon_0+i\,\sin\theta(r)\,\gamma_*\,\chi^\pm_{S^2}\,\otimes\sigma^3 \varepsilon_0  \right)\ ,
 \label{killingMkW3S2}
 \end{split}
\end{equation}
where $\chi^\pm_{S^2}$ satify the equations \eqref{killingS2}. By further imposing the algebraic condition
\begin{equation}
 (\gamma_*\otimes \sigma^1)\,(\chi^\pm_{S^2}\otimes \varepsilon_0)=\pm\,\chi^\pm_{S^2}\otimes \varepsilon_0\ ,
 \label{projMkW3S2}
\end{equation}
the BPS equations for the background \eqref{MkW3S2ansatz} take the form
\begin{equation}
 \begin{split}
   U^\prime=&\ -\frac{1}{2}\,e^{V}\,\cos(2\theta)^{-1}\left((3+\cos(4\theta))\,f+2 \sin{(2\theta)}^2\,X\,D_Xf   \right)\ ,\\
  W^\prime=&\ \frac{1}{2}\,e^{V}\,\cos(2\theta)^{-1}\left((-5+\cos(4\theta))\,f+2 \sin{(2\theta)}^2\,X\,D_Xf   \right)\ ,\\
  b^\prime=&\ \frac{4}{X^2} \,e^{V+2 W}\, \sin (2 \theta )\,\left(f-\,X\,D_Xf     \right)\ ,\\
  X^\prime=&\ \frac12\, e^{V}\,\cos(2\theta)^{-1}\, X\, \left(2\,\sin(2\theta)^2\,f+ (3+\cos(4\theta))\,X\,D_Xf   \right)\ ,\\
  Y^\prime=&\ -\frac{Y}{4}\,e^{V}\,\cos(2\theta)^{-1}\left((3+\cos(4\theta))\,f+2 \sin{(2\theta)}^2\,X\,D_Xf   \right)\ ,\\
  \theta^\prime=&\ e^{V}  \,\sin (2 \theta )\,\left(f-X\,D_Xf     \right)\ .
\label{floweqMkW3S2}
  \end{split}
\end{equation}
Just as in the previous examples we have two additional constraints
\begin{equation}
 \begin{split}
  b\overset{!}{=}&\ -\frac{4}{m} \,e^{2 W}\, \tan (2 \theta )\,X\,\left(f+X\,D_Xf\right)\ ,\\
  R\overset{!}{=}&\ 2\, e^{W}\, \tan (2 \theta )\,\left(3\,f+X\,D_Xf\right)\ ,
  \label{constraintsMkw3S2}
 \end{split}
\end{equation}
which are automatically satified if $f$ has the form of the prepotential \eqref{superpotential}. The gauge choice
\begin{equation}
 e^{V}=\left(\sin{(2\theta)}\,(f-X\,D_Xf)  \right)^{-1}
 \label{gaugechoiceMkW3S2}
\end{equation}
restricts the range of the $r$ coordinate to $(0,\frac{\pi}{4})$. Thanks to the choice in \eqref{gaugechoiceMkW3S2},
we can integrate \eqref{floweqMkW3S2} to obtain the following solution
\begin{equation}
 \begin{split}
  e^{2U}=&\ \left(2-\cos(4 r)\right)^{1/2}\,\sin(2r)^{-2}\ ,\\
  e^{2W}=&\ \left(2-\cos(4 r)\right)^{1/2}\,\tan(2r)^{-2}\ ,\\
  e^{2V}=&\ \frac{2^{5/4}\,3^{3/2}}{m^{1/2}\,g^{3/2}}\, \left(2-\cos(4 r)\right)^{-3/2}\,\sin(2r)^{-2}\ ,\\
  b=&\ -\frac{2^{5/4}\,g^{1/2}}{\,3^{1/2}\,m^{1/2}}\cos(2r)^{2}\, \tan (2 r)^{-2}\ ,\\
  X=&\ \frac{3^{1/4}\,m^{1/4}}{2^{1/8}\,g^{1/4}} \,\left(2-\cos(4 r)\right)^{-1/4}\ ,\\
  Y=&\ \left(2-\cos(4 r)\right)^{1/8}\,\sin(2r)^{-1/2}\ ,\\
  \theta=&\ r\ .
   \label{flowMkW3S2}
 \end{split}
\end{equation}
From the constraints \eqref{constraintsMkw3S2} we obtain the expression for the inverse of the radius of the 2-sphere
\begin{equation}
 R=2^{3/8}\,3^{1/4}\,(g^{3}\,m)^{1/4}\ ,
 \label{RMkW3S2}
\end{equation}
for $g>0$ and $m>0$. Imposing \eqref{RMkW3S2} the equations of motion \eqref{eom} are satified by the flow \eqref{flowMkW3S2}.
In the limit $r \rightarrow 0$ the background \eqref{flowAdS3R2} reproduces locally the $\mrm{AdS}_6$ vacuum  \eqref{AdS6}, while in the limit $r \rightarrow \frac{\pi}{4}$ the solution is singular.

\subsection{Background with $M_3=\mrm{AdS}_3$ and $\Sigma_2=S^{2}$}
\label{sec:AdS3S2}

Let's now move to the most involved case where $M_3=\mrm{AdS}_3$ and $\Sigma_2=S^{2}$. In this case the 6d background takes the following form
\begin{equation}
 \begin{split}
  ds_6^2=&\ e^{2U(r)}\,ds^2_{{\scriptsize \mrm{AdS}_3}}+e^{2V(r)}\,dr^2+e^{2W(r)}\,ds^2_{S^2}\ ,\\
  \ma{B}_{(2)}=&\ b(r)\,\text{vol}_{S^2}\ ,\\
  X=&\ X(r)\ .
  \label{AdS33S2ansatz}
  \end{split}
\end{equation}
We take a Killing spinor of the following form
\begin{equation}
\begin{split}
 \zeta(r)=&\ \zeta^++i\,B\,\Gamma_*\,\zeta^-\ ,\\
 \zeta^\pm=&\ Y(r)\,\eta_{{\scriptsize \mrm{AdS}_3}}\,\otimes \,\left(\cos\theta(r)\,\chi^\pm_{S^2}\otimes \varepsilon_0+i\,\sin\theta(r)\,\gamma_*\,\chi^\pm_{S^2}\,\otimes\sigma^3 \varepsilon_0  \right)\ ,
 \label{killingAdS3S2}
 \end{split}
\end{equation}
where $\eta_{{\scriptsize \mrm{AdS}_3}}$ and $\chi^\pm_{S^2}$ respectively satisfy the Killing spinor equations \eqref{killingM3} and \eqref{killingS2}.
As in section \ref{sec:AdS3R2}, in order to simplify the derivation of the first-order flow equations, we parametrize the $\mrm{AdS}_3$ foliation with the Hopf coordinates \eqref{AdS3} since this is equivalent to replacing $\eta_{{\scriptsize \mrm{AdS}_3}}$ by $\eta_{\mathbb{R}^{1,2}}$ inside \eqref{killingAdS3S2}.

An explicit realization of \eqref{AdS33S2ansatz} is defined by a specific relation between $R$ and $L$ characterizing the geometry of the 6d background. In this section we derive two solutions corresponding to two different relations between $R$ and $L$. 

Let's start with the simplest case with two equal warp factors in \eqref{AdS33S2ansatz}, \emph{i.e.} $U(r)=W(r)$. If one imposes the algebraic conditions \eqref{projMkW3S2} on \eqref{killingAdS3S2}, the SUSY variations \eqref{SUSYvariation6d} imply a non-trivial set of BPS equations if and only if
\begin{equation}
 R=2L\qquad \text{and} \qquad \theta(r)=0\ .
\end{equation}
We obtain the following set of first-order equations
\begin{equation}
 \begin{split}
  U^\prime=&\ -2\,e^{V}\,f\,,\qquad \ Y^\prime=-Y\,e^{V}\,f\ ,\\
  b^\prime=&\ \frac{2\,e^{U+V}\,L}{X^2}\,,\qquad X^\prime=2\,e^{V}\,X^2\,D_Xf\ ,
  \label{chargedDW}
 \end{split}
\end{equation}
with the constraint
\begin{equation}
 b\overset{!}{=}-\frac{2\,e^{U}\,X\,L}{m}\ .
\end{equation}
The above expression is compatible with the BPS flow equations in \eqref{chargedDW} if $f$ is given by \eqref{superpotential}. If we further choose
\begin{equation}
 e^V=\left(2\,X^2\,D_Xf\right)^{-1}\ ,
\end{equation}
we can integrate the system in \eqref{chargedDW} to obtain
\begin{equation}
 \begin{split}
  e^{2U}=&\  \left(\frac{r}{\sqrt{2}\, g \,r^4-3 \,m}\right)^{2/3}\ ,\\
   e^{2V}=&\  \left(\frac{4\,r^2}{\sqrt{2}\, g \,r^4-3 \,m}\right)^{2}\ ,\\
   Y=&\ \left(\frac{r}{\sqrt{2}\, g \,r^4-3 \,m}\right)^{1/6}\ ,\\
   b=&\ -\frac{2\,r^{4/3}\,L}{m\,(\sqrt2\,g\,r^4-3\,m)^{1/3}}\ ,\\
   X=&\ r\ ,
   \label{chargedDWsol}
 \end{split}
\end{equation}
with $r$ running between 0 and 1 if we choose $m$ and $g$ such that \eqref{m&guplift} holds.
We point out that the $\mathrm{AdS}_3$ slicing is responsible for the non-trivial profile of the 2-form. In a sense, the flow \eqref{chargedDWsol} is the ``charged'' generalization of the domain wall solution \eqref{DWsolution}.
In the $r\rightarrow 1^{-}$ limit, the solution \eqref{chargedDWsol} locally reproduces the $\mrm{AdS}_6$ vacuum \eqref{AdS6} with $m=\frac{\sqrt2\,g}{3}$, while in $r\rightarrow 0^{+}$ it manifests a singular behavior.

Let us now consider the most general case of backgrounds of the form \eqref{AdS33S2ansatz} with two independent warp factors. Given the Killing spinor \eqref{killingAdS3S2} satisfying the algebraic conditions \eqref{projMkW3S2}, we obtain a set of BPS equations of the form
\begin{equation}
 \begin{split}
  U^\prime=&-\frac{1}{2}\,e^{V}\,\cos(2\theta)^{-1}\left((3+\cos(4\theta))\,f+2 \sin{(2\theta)}^2\,X\,D_Xf+L\,e^{-U}\sin(2\theta)   \right)\ ,\\
  W^\prime=&\frac{1}{2}\,e^{V}\,\cos(2\theta)^{-1}\left((-5+\cos(4\theta))\,f+2 \sin{(2\theta)}^2\,X\,D_Xf  -L\,e^{-U}\sin(2\theta)  \right)\ ,\\
  b^\prime=&\frac{2\,e^{V+2 W}}{X^2}\,\left(L\,e^{-U}+2 \sin (2 \theta )\,\left(f-\,X\,D_Xf     \right)\right)\ ,\\
  \theta^\prime=& e^{V}  \,\sin (2 \theta )\,\left(f-X\,D_Xf     \right)\ ,\\
  Y^\prime=&-\frac{Y}{4}\,e^{V}\,\cos(2\theta)^{-1}\left((3+\cos(4\theta))\,f+2 \sin{(2\theta)}^2\,X\,D_Xf+L\,e^{-U}\sin(2\theta)   \right)\ ,\\
  X^\prime=&\frac12\, e^{V}\, X\,\left(L\,e^{-U}\tan(2\theta)+\cos(2\theta)^{-1}\, \left(2\,\sin(2\theta)^2\,f+ (3+\cos(4\theta))\,X\,D_Xf   \right)\right)\ .
\label{floweqAdS3S2}
  \end{split}
\end{equation}
The equations \eqref{floweqAdS3S2} have to be supplemented with the constraints
\begin{equation}
 \begin{split}
  b\overset{!}{=}&\ -\frac{2\,e^{2 W}}{m}\,X\,\cos (2 \theta )^{-1}\,\left(L\,e^{-U}+2\,\sin(2\theta)\,\left(f+\,X\,D_Xf     \right)\right)\ ,\\
  R\overset{!}{=}&\ 2\,e^{-U+W}\,L\,\cos(2\theta)^{-1}+ 2\, e^{W}\, \tan (2 \theta )\,\left(3\,f+X\,D_Xf\right)\ ,
  \label{constraintsAdS3S2}
 \end{split}
\end{equation}
which are automatically satified if $f$ has the form of the prepotential \eqref{superpotential}.
If we choose
\begin{equation}
 e^{V}=\left(\sin{(2\theta)}\,(f-X\,D_Xf)  \right)^{-1}\ ,
 \label{gaugechoiceAdS3S2}
\end{equation}
the solution of \eqref{floweqAdS3S2} is given by
\begin{equation}
 \begin{split}
  e^{2U}=&\ \frac{2^{1/4}\,g^{1/2}}{3^{1/2}\,m^{1/2}}\,\sin{(2r)}^{-1}\,\left(\sin(2 r)^{-2}+6\right)^{1/2}\ ,\\
  e^{2W}=&\ 2^{-5/4}\,(-g)^{1/2}\,\tan(2r)^{-2}\,\left(4-3\cos(4 r)\right)^{1/2}\ ,\\
  e^{2V}=&\ \frac{2^{5/4}\,3^{3/2}}{m^{1/2}\,g^{3/2}}\,\frac{ \left(\sin(2 r)^{-2}+6\right)^{1/2}\,\sin(2r)^{-1}}{\left(4-3\cos(4 r)\right)^{2}}\ ,\\
  b=&\ -\frac{3^{1/2}\,4\,g}{(-m)^{1/2}}\cos(2r)^{2}\, \tan (2 r)^{-2}\ ,\\
  X=&\ \frac{3^{1/4}\,m^{1/4}}{2^{1/8}\,g^{1/4}} \,\left(4-3\,\cos(4 r)\right)^{-1/4}\ ,\\
  Y=&\ \frac{2^{1/16}\,g^{1/8}}{3^{1/8}\,m^{1/8}}\,\sin{(2r)}^{-1/4}\,\left(\sin(2 r)^{-2}+6\right)^{1/8}\ ,\\
  \theta=&\ r\ ,
   \label{flowAdS3S2}
 \end{split}
\end{equation}
with $r$ varying between 0 and $\frac{\pi}{4}$.
The above solution solves the equations of motion \eqref{eom} provided that
\begin{equation}
 L=\frac{2\sqrt2\,g}{3}\qquad \text{and}\qquad R=\frac{2^{3/4}\,7}{3^{3/4}}\,g\,(-m)^{-1/4}\ ,
\label{L&R}
\end{equation}
with $m<0$ and $g<0$. The values \eqref{L&R} satisfies \eqref{constraintsAdS3S2} that, if evaluated on the solution, takes the following form
\begin{equation}
 R=(-6\,m)^{1/4}\,\left(2\,L+\sqrt2\,g \right)\,.
 \label{onshellconstraint}
\end{equation}
In the $r\rightarrow 0^{+}$ limit the flow \eqref{flowAdS3S2} reproduces locally the $\mrm{AdS}_6$ vacuum \eqref{AdS6}. The IR regime, \emph{i.e.} $r\rightarrow \left(\frac{\pi}{4}\right)^{-}$ is particularly interesting since the scalar potential is finite and the flows turns out to be locally described by $\mrm{AdS}_3\times T^3$ with
\begin{equation}
X=\frac{3^{1/4}\,m^{1/4}}{2^{1/8}\,7^{1/4}\,g^{1/4}}\qquad \text{and} \qquad \ma F_{(3)\,345}=2^{13/8}\,3^{3/4}\,7^{1/4}\,g^{5/4}\,m^{-1/4}\ ,
\end{equation}
where we point out that the particular relation \eqref{onshellconstraint} between $R$ and $L$ turns out to be crucial to reproduce the $\mrm{AdS}_3\times T^3$ geometry.

\section{Surface Defects within the $\ma N=2$ $\mrm{SCFT}_5$}

In section~\ref{D4D8system} we briefly reviewed the main features of $\mrm{AdS}_6/\mrm{CFT}_5$ in massive IIA string theory. Our present goal is now that of providing a 10d interpretation for the $\mrm{AdS}_3$ slicing characterizing the 6d backgrounds obtained in section \ref{sec:AdS3S2}. For the sake of simplicity we will consider the charged domain wall \eqref{chargedDWsol}. This flow is a 6d warped product $\mrm{AdS}_3\times \ma M_3$ where $\ma M_3$ is given by a 2-sphere fibered over an interval. Asymptotically ($r\rightarrow 1^{-}$), the solution locally reproduces $\mathrm{AdS}_6$, while in the IR ($r\rightarrow 0^{+}$) it possesses a singularity. We claim that this divergent behavior is related to the intersection of the D4-D8 system, originating the $\mrm{AdS}_6$ vacuum, with a bound state of D2-NS5-D6 defect branes. The $\mrm{AdS}_3$ slicing captures exactly the low-energy regime of this intersection. In particular, the presence of the 2-form gauge potential, whose field strength $\ma F_{(3)}$ fills the transverse space $\ma M_3$, realizes a partial symmetry breaking within the $\mrm{AdS}_6$ vacuum. In this way the divergent behavior appearing in the IR limit describes the regime in which we get infinitely close to the defect branes, namely the D2-NS5-D6.

From the point of view of the dual field theory, this phenomenon is well encoded by a $\ma N=(0,4)$ $\mrm{SCFT}_2$ living on a surface conformal defect \cite{Karch:2000gx} within the $\ma N=2$ $\mrm{SCFT}_5$ dual to $\mrm{AdS}_6$. This defect field theory can be seen as a ``position dependent'' deformation \cite{Clark:2004sb} of the $\mrm{SCFT}_5$ partially breaking the $\mathrm{SO}(2,5)$ conformal invariance in the 5d bulk, while still keeping intact only those conformal isometries allowing non-trivial boundary conditions between the D4-D8 system and the defect branes.

In this section we will firstly consider in more detail the 6d solution \eqref{chargedDWsol} and its uplift to massive IIA using the formulas \eqref{truncationansatz} and \eqref{10dfluxes}. We will then consider the 10d background corresponding to the bound state D2-D4-NS5-D6-D8 realizing the $\mrm{AdS}_3$ slicing and we will propose a holographic interpretation of our $\mrm{AdS}_3\times \ma M_3$ background as a conformal defect within the $\ma N=2$ $\mrm{SCFT}_5$.

\subsection{Charged Domain Wall and Massive IIA Uplift}

Let us go back to the explicit form of the background \eqref{chargedDW}. The line element is given by
\begin{equation}
 \begin{split}
  ds_6^2=&\ e^{2U(r)}\left( \,ds^2_{{\scriptsize \mrm{AdS}_3}}+ds^2_{S^2}\right)+e^{2V(r)}\,dr^2\ ,\\
  \ma{B}_{(2)}=&\ b(r)\,\text{vol}_{S^2}\ ,\\
   X=&\ X(r)\ ,
  \label{CDWansatz}
  \end{split}
\end{equation}
where
\begin{equation}
 \begin{split}
  e^{2U}= &\ \frac{2^{-1/3}}{g^{2/3}}\,\left(\frac{r}{r^4-1}\right)^{2/3}\ , \qquad e^{2V}=\frac{8}{g^2}\, 
  \frac{r^4}{\left( r^4-1\right)^2}\ ,\\
   b=&\ -\frac{2^{1/3}\,3}{g^{4/3}}\,\frac{L\,r^{4/3}}{(r^4-1)^{1/3}}\ ,\qquad \ X=r\ ,
   \label{CDW}
 \end{split}
\end{equation}
where we made the choice \eqref{m&guplift} on $m$ and with $r$ running between 0 and 1. Let's consider more in detail the UV and IR regimes. As $r \rightarrow 1^{-}$ one obtains
\begin{equation}
 \begin{split}
  \mathcal{R}_{6}=&\ -\frac{20}{3}\,g^2+\ma O (1-r)^{2/3}\ ,\\
  X=&\ 1+\ma O (1-r)\ ,
  \label{UVCDW}
 \end{split}
\end{equation}
where $\mathcal{R}_{6}$ is the scalar curvature. The asymptotic background \eqref{UVCDW} reproduces only locally $\mrm{AdS}_6$ and this is mainly due to the presence of the running 2-form. The $\mrm{AdS}_6$ vacuum emerges in the asymptotics only as a leading local effect, but globally a 2-form charge is still present. As for the $r\rightarrow 0^{+}$ limit, one finds that
\begin{equation}
 \begin{split}
  e^{2U}=&\ \frac{g^{-2/3}}{2^{2/3}}\,r^{2/3}+\ma O(r^{7/3})\ ,\qquad \ e^{2V}=\frac{8\,r^4}{g^2}+\ma O(r^5)\ ,\\
  b=&\ \frac{2^{1/3}\,3\,L}{g^{4/3}}\,r^{4/3}+\ma O(r^{7/3})\ ,\qquad X=r+\ma O(r^5)\ .
   \label{IRCDW}
 \end{split}
\end{equation}
In this regime the background \eqref{IRCDW} is manifestly singular, so then we want to study the uplift to massive IIA supergravity to shed some light on the origin of this divergent behavior. If we plug the 6d bacgkround \eqref{CDWansatz} into the uplift formulas \cite{Cvetic:1999un} \eqref{truncationansatz} and \eqref{10dfluxes}, the 10d metric has the form
\begin{equation}
 \begin{split}
  ds^2_{10}=s^{-1/3}\,X(r)^{-1/2}\,\Delta^{1/2}\,\left[e^{2U(r)}\left( \,ds^2_{{\scriptsize \mrm{AdS}_3}}+ds^2_{S^2}\right)+e^{2V(r)}\,dr^2+2g^{-2}\,X(r)^{2}\,ds_{{\scriptsize{\tilde{S}^4}}}^2\right]\ ,
  \label{upliftCDW}
 \end{split}
\end{equation}
where $\Delta=X(r)\,c^2+X(r)^{-3}\,s^2$ and $ds_{\tilde{S}^4}^2$ is the metric of a squashed 4-sphere 
\begin{equation}
ds_{{\scriptsize{\tilde{S}^4}}}^2=d\xi^2+\frac14\,\Delta^{-1}\,X(r)^{-3}\,c^{2}\,ds_{S^3}^2\ ,
 \label{intspaceCDW}
\end{equation}
with $c=\cos\xi$ and $s=\sin \xi$ and $ds_{S^3}^2$ is the metric of a round\footnote{The 3-sphere is round because the vectors $A^i$ vanish for the charged domain wall.} $S^3$. The 10d fluxes have the form
\begin{equation}
\label{10dfluxesCDW}
 \begin{split}
 F_{(4)}&=-\frac{\sqrt{2}}{6} g^{-3}\,\Delta^{-2}\,c^3\,s^{1/3}\left[ U\,d\xi+6s\,c\,X(r)^{-3}\,X^\prime(r)\,dr\right]\,\wedge\,\text{vol}_{S^3}\\
  &+e^{-2U(r)-V(r)}\,s^{1/3}\,c\left[\sqrt2 \,g^{-1}\,X(r)^4\,b^\prime(r)\,d\xi-m\,s\,c^{-1}\,X(r)^{-2}\,e^{2V(r)} b(r)\, dr\right]\wedge \text{vol}_{{\scriptsize \mrm{AdS}_3}} \ ,\\
  F_{(2)}&=m\,s^{2/3}\,b(r)\,\text{vol}_{S^2}\ ,\qquad \quad\ H_{(3)}=s^{2/3}\left[b^\prime(r)\,dr +\sqrt2\,m\,g^{-1}\,s\,c\,b(r)\,d\xi\right]\wedge\text{vol}_{S^2}\ ,\\
  e^{\Phi}&=s^{-5/6}\,\Delta^{1/4}\,X(r)^{-5/4}\ ,\qquad F_{(0)}=m\ .
 \end{split}
\end{equation}
where $U=X(r)^{-6}\,s^2-3X(r)^2\,c^2+4\,X(r)^{-2}\,c^2-6\,X(r)^{-2}$, while $\text{vol}_{S^3}$ and $\text{vol}_{S^2}$ are respectively the volume form of the internal $S^3$ included in \eqref{intspaceCDW} and the volume form of the 2-sphere appearing into the 6d bacgkround.

The background \eqref{upliftCDW} with fluxes \eqref{10dfluxesCDW} is a solution of massive IIA supergravity describing a warped geometry of the type $\mrm{AdS}_3\times S^2 \times S^3$ fibered over two intervals $I_r\times I_\xi$. In the same way as this solution, also the other flows of section~\ref{flows} admit similar uplifts to 10d. In particular, the 10d background corresponding to the solution \eqref{flowAdS3S2}, in the $r\rightarrow 0$ limit, is locally described by the warped geometry $\mrm{AdS}_3\times T^3 \times \tilde{S}^4$ where $\tilde{S}^4$ is a fibration of a 3-sphere over the interval $I_\xi$.

\subsection{Defect $\mrm{SCFT}_2$ and the $\mrm{AdS}_3\times S^2\times S^3\times I^2$ Solution}

Let us now address the interpretation of the charged domain wall \eqref{chargedDW} in terms of physics of branes in massive type IIA string theory. In section \ref{D4D8system} we reviewed the main properties of the 10d solution \eqref{BOsolution} describing the low-energy regime of a D4-D8 system. We saw that the near-horizon limit is described by the vacuum geometry $\mrm{AdS}_6\times S^4$ from which a clear holographic interpretation in terms of a $d=5$ $\ma N=2$ $\mrm{SCFT}_5$ comes out.

We can now look at this $\mrm{SCFT}_5$ as a ``mother'' CFT whose conformal invariance is partially broken by a deformation associated with a position dependent coupling produced by D2-NS5-D6 branes ending on the bound state D4-D8. The low-energy description of this intersection is realized by the emergence of the warped background $\mrm{AdS}_3\times \ma M_3$ that partially breaks the isometries of the $\mrm{AdS}_6$ vacuum.

\begin{table}[h!]
\renewcommand{\arraystretch}{1}
\begin{center}
\scalebox{1}[1]{
\begin{tabular}{c||c c|c c c c|c||c c c}
branes & $t$ & $y$ & $\rho$ & $\varphi^{1}$ & $\varphi^{2}$ & $\varphi^{3}$ & $z$ & $r$ & $\theta^{1}$ & $\theta^{2}$ \\
\hline \hline
$\mrm{D}8$ & $\times$ & $\times$ & $\times$ & $\times$ & $\times$ & $\times$ & $-$ & $\times$ & $\times$ & $\times$ \\

$\mrm{D}4$ & $\times$ & $\times$ & $-$ & $-$ & $-$ & $-$ & $-$ & $\times$ & $\times$ & $\times$ \\
\hline
$\mrm{D}2$ & $\times$ & $\times$ & $-$ & $-$ & $-$ & $-$ & $\times$ & $-$ & $-$ & $-$ \\

$\mrm{NS}5$ & $\times$ & $\times$ & $\times$ & $\times$ & $\times$ & $\times$ & $-$ & $-$ & $-$ & $-$ \\
$\mrm{D}6$ & $\times$ & $\times$ & $\times$ & $\times$ & $\times$ & $\times$ & $\times$ & $-$ & $-$ & $-$ \\

\end{tabular}
}
\end{center}
\caption{{\it The brane picture underlying the $\ma N=(0,4)$ $\mrm{SCFT}_2$ defect theory described by D2-NS5-D6 branes ending on an D4-D8 intersection. The system is $\mrm{BPS}/8$.}} \label{Table:branes}
\end{table}
The 10d solution describing a D2-D4-NS5-D6-D8 system has been obtained in \cite{Dibitetto:2017klx} generalizing the corresponding massless background originally found in \cite{Boonstra:1998yu}. The solution is an example of ``non-standard'' intersection since the are no common transverse directions and the only solution that could be obtained by applying the standard harmonic superposition principle would be 10d flat space, \emph{i.e.} all $H$ functions equal to $1$. The explicit form of this non-standard solution is given by \cite{Dibitetto:2017klx}
\begin{equation}
\label{brane_sol}
\begin{split}
\D s_{10}^{2}  = &\ S^{-1/2}H_{\scriptsize{\mrm{D}2}}^{-1/2}H_{\scriptsize{\mrm{D}4}}^{-1/2}\,d s_{\scriptsize{\mathbb{R}^{1,1}}}^{2}\,+\, 
S^{-1/2}H_{\scriptsize{\mrm{D}2}}^{1/2}H_{\scriptsize{\mrm{D}4}}^{1/2}\,\left(d \rho^{2}+\rho^{2}\,d s_{\scriptsize{S^3}}^{2}\right) \,\\
 &+  K\,S^{-1/2}H_{\scriptsize{\mrm{D}2}}^{-1/2}H_{\scriptsize{\mrm{D}4}}^{1/2}\,d z^{2}\,+\,K\,S^{1/2}H_{\scriptsize{\mrm{D}2}}^{1/2}H_{\scriptsize{\mrm{D}4}}^{-1/2}\,\left(d r^{2}+r^{2}\,d s_{\scriptsize{S^2}}^{2}\right) \ ,\\[2mm]
e^{\Phi} = &\ g_{s}\,K^{1/2}\,S^{-3/4}H_{\scriptsize{\mrm{D}2}}^{1/4}H_{\scriptsize{\mrm{D}4}}^{-1/4}  \ ,\\
H_{(3)}  = &\ \frac{\partial}{\partial z}\left(KS\right)\mrm{vol}_{S^3} \,-\,dz\,\wedge\,\star_{3}\, \D K  \ ,\\[1mm]
F_{(0)}  = &\ m  \ ,\\
F_{(2)}  = &\ -g_{s}^{-1}\,\star_{3}\, \D S \ ,\\
F_{(4)}  = &\ g_{s}^{-1}\,\mrm{vol}_{\mathbb{R}^{1,1}}\,\wedge\,d z\,\wedge\, \D H_{\scriptsize{\mrm{D}2}}^{-1} \, + \, 
\star_{10}\left(\textrm{vol}_{\mathbb{R}^{1,1}}\,\wedge\,\mrm{vol}_{S^3}\wedge\, H_{\scriptsize{\mrm{D}4}}^{-1}\right)  \ .
\end{split}
\end{equation}
The two functions $K(z,r)$ and $S(z,r)$ satsify the equations \cite{Imamura:2001cr}
\begin{equation}
\begin{split}
mg_{s}\,K \,-\, \frac{\partial S}{\partial z}  = & \ 0 \ , \\
\Delta_{(3)}S \, + \, \frac{1}{2}\frac{\partial^{2}}{\partial z^{2}} S^{2}  = &\ 0 \ .
\end{split}
\end{equation}
These relations must hold in order to satify equations of motion \eqref{massiveIIAeoms} and Bianchi identities \eqref{massiveIIAbianchi} and their explicit solutions turn out to describe a rich plethora of different physical sytems.
The background \eqref{BOsolution2} and the corresponding $\mathrm{AdS}_6\times S^4$ vacuum in the near-horizon can be found by choosing
\begin{equation}
\begin{split}
 S=&\ (2mg_s\,z)^{1/2}\ ,\qquad\, K\ =\ (2mg_s\,z)^{-1/2}\ , \\
H_{\scriptsize{\mrm{D}2}}=&\ 1\ , \qquad\qquad\qquad H_{\scriptsize{\mrm{D}4}}=\ 1+\frac{Q_{\mathrm{D}4}}{\left(\rho^{2}+\frac{4}{9}\frac{(2z)^{3/2}}{(g_s m)^{1/2}}\right)^{5/3}}\ ,
\end{split} 
 \label{SKeq}
\end{equation}
and by subsequently performing the change of coordinates $z \rightarrow \frac{m\,g_s}{2}\,z^2$.
As we can notice from \eqref{SKeq} physics described by the solution \eqref{brane_sol} depends explicitly on the choice of particular solutions for $S$ and $K$.
As we showed in \cite{Dibitetto:2017klx}, from \eqref{brane_sol} it is also possible to find the massive IIA $\mathrm{AdS}_7\times \tilde{S}^3$ vacuum \cite{Apruzzi:2013yva} that, in turn, can be obtained as a vacuum of the $\ma N=1$ minimal gauged supergravity in $d=7$ thanks to the consistent truncation from massive IIA over a squashed 3-sphere \cite{Passias:2015gya}.
Within \eqref{brane_sol} a warped solution $\mrm{AdS}_3\times S^3\times S^2\times I^2$, with $I^2$ describing two intervals on which the $S^2$ and the $S^3$ are respectively fibered, has been derived in the near-horizon of \eqref{brane_sol} by choosing
\begin{equation}
\begin{split}
H_{\scriptsize{\mrm{D}2}}(\rho,r) = &\ \left(1+\frac{Q_{\scriptsize{\mrm{D}4}}}{\rho^{2}}\right)\left(1+\frac{Q_{\scriptsize{\mrm{D}6}}}{r}\right)\ ,\\
H_{\scriptsize{\mrm{D}4}}(\rho) = &\ \left(1+\frac{Q_{\scriptsize{\mrm{D}4}}}{\rho^{2}}\right) \  ,
\label{H2H4}
\end{split}
\end{equation}
and for suitable expressions of $S$ and $K$. This $\mrm{AdS}_3\times S^3\times S^2\times I^2$ near-horizon is captured by a warped background $\mathrm{AdS}_3\times S^3\times I_{r'}$ describing\footnote{For sake of clarity, the coordinates associated with the 7d flow will be called as $r', \xi',\dots$.} a charged domain wall in 7d minimal $\ma N=1$ gauged supergravity with a running 3-form gauge potential and the dilaton \cite{Dibitetto:2017klx}.

The very interesting fact we point out is that the above $\mrm{AdS}_3$ near-horizon has the same structure of fluxes and 10d metric of our 6d background uplifted to massiva IIA \eqref{upliftCDW} and it also preserves the same amount of SUSY.
The unique difference between \eqref{upliftCDW} and (3.13) of \cite{Dibitetto:2017klx} is in the parametrization of the 10d background. In our case the $S^2$ is related to the 6d background and the $S^3$ is associated to the internal squashed 4-sphere $I_\xi\times S^3$, \emph{i.e.} we have $\mrm{AdS}_3\times S^2\times S^3\times I^2$, while the 7d case is exactly specular, \emph{i.e.} the squashed 3-sphere $I_{\xi'}\times S^2$ defines the truncation and the near horizon can be written as $\mrm{AdS}_3\times S^3\times S^2\times I^2$. In other words, we could say that the coordinates $(r, \xi)$ of the uplift \eqref{upliftCDW} exchange their role when we look at the 10d background split as a 7+3 rather than a 6+4 manifold.

The 10d solution \eqref{brane_sol}, \eqref{H2H4} realizes the brane picture of \eqref{CDWansatz} and then, we may holographically interpret our 6d warped background \eqref{CDW} as a conformal defect within the $\ma N=2$ $\mrm{SCFT}_5$. The defect is realized by a $\ma N=(0,4)$ $\mrm{SCFT}_2$ dual to the $\mrm{AdS}_3$ foliation and it breaks the $\mathrm{SO}(5,2)$ conformal group of the 5d mother SCFT through a relavant deformation driven by a position-dependent coupling.

The above arguments provide compelling evidence to infer that the defect $\mrm{SCFT}_2$ is actually the same as the one obtained in \cite{Dibitetto:2017klx}. We remind that, in the latter case, a warped $\mrm{AdS}_3\times S^3\times I_{r'}$ solution within 7d minimal $\ma N=1$ supergravity was capturing the physics of D2-D4 branes intersecting the bound state NS5-D6-D8 and giving rise to a surface defect within the $\ma N=(1,0)$ $\mrm{SCFT}_6$. In our actual case we have a $\mrm{AdS}_3\times S^2\times I_r$ 
warped flow associated to D2-NS5-D6 ending on the D4-D8 system. In both cases we have a $p$-form gauge potential which makes the existence of such an $\mrm{AdS}_3$ slicing possible.  Zooming on the defect, we obtain the same $\mrm{AdS}_3$ solution, up to a change of parametrization exchanging the roles of the coordinates within the intervals in $I^2$ and thus swapping $S^3$ and $S^2$.

Finally we point out that these arguments hint at the existence of a deeper relation between the two lower-dimensional supergravities giving rise to these $\mrm{AdS}_3$ warped backgrounds.
The existence of a possible link between them was already adressed in \cite{Jeong:2013jfc} within the slightly different context of non-Abelian T-duality and the possibility of uplifting the Romans' theory to type IIB. Our conjecture here is about the existence of a 3d $\ma N=4$ gauged supergravity realizing a consistent truncation respectively of $F(4)$ gauged supergravity over a squashed 3-sphere or, alternatively, of 7d minimal $\ma N=1$ gauged supergravity over a squashed 4-sphere. As a consequence, this 3d gauged supergravity should include an $\mrm{AdS}_3$ vacuum capturing the IR physics of surface defects of both brane systems, \emph{i.e.} D4-D8 and NS5-D6-D8.

\subsection{One-Point Correlation Funtions}

In conclusion, we want to provide a holographic test in support of the presence of a $\ma N=(0,4)$ $\mrm{SCFT}_2$ defect theory. As we already mentioned, the coupling of the bulk theory to the defect induces the breaking of the 5d conformal group $\mathrm{SO}(2,5)$. This automatically implies that, in this case, the 1-point functions of the 5d ``mother'' field theory are no longer vanishing. Such a fact stems at leading order from non-vanishing defect to bulk correlators. By making use of the standard holographic dictionary \cite{Clark:2004sb}, we can sketch the derivation in two different ways and see explicitly that the resulting position-dependence of the coupling in the 5d theory realizing the defect matches.

Let's consider the extrapolation from the bulk side. The boundary of our 6d background \eqref{CDWansatz} is located at $r=1$. The metric can be rewritten as
\begin{equation}
\begin{split}
 ds^2_6=&\ F^{-2}\left(ds^2_{\mathbb{R}^{1,5}}+\rho^2\,dR^2\right)\ ,\\
 ds^2_{\mathbb{R}^{1,5}}=&\ ds^2_{\mathbb{R}^{1,1}}+d\rho^2+\rho^2\,ds_{S^2}^2\ ,
 \end{split}
\end{equation}
with $F=\rho\,e^{-U}$ and $dR=e^{V-U}\,dr$. The coordinate $\rho$ is the $\mrm{AdS}_3$ radial coordinate and it fixes the location of the defect at $\rho=0$.

The idea is to view the scalar $X$ as the bulk field associated with the deformation induced by the defect. Its normalized mass at the boundary is given by \cite{Andrianopoli:2001rs}
\begin{equation}
 m_X^2=-6=\Delta_X(\Delta_X-5)\ ,
\end{equation}
whence $\Delta_X=3$.
If we consider the asymptotic behavior of $X$ given in \eqref{UVCDW}, we can cast it as a function of $R$ as
\begin{equation}
 X(R)=1-\frac{g^2}{3^3 2^5}\,R^3\ .
\end{equation}
As usual in holography, the vev of $X$ is associated to the 1-point function of the dual operator $\ma O_X$ as it follows
\begin{equation}
 X=1-b\,\langle \ma O_X \rangle \,F^{\Delta_X}+\dots\ .
\end{equation}
Finally by comparing the last two relations we obtain
\begin{equation}
 \langle \ma O_X \rangle =\frac{1}{\sqrt2\,g\,b}\rho^{-3}\ .
 \label{1pointbulk}
\end{equation}
From the $\mrm{SCFT}_5$ side,  we can proceed through conformal perturbation techniques. We interpret the defect as a running vev written in terms of a position-dependent coupling $\phi(\rho)$ and producing a deformation of the type $\gamma\,\phi(\rho)\,\ma O_X$, where $\gamma$ is a dimensionsless coupling associated with the anomalous dimension of $\ma O_X$.
We can treat this deformation as a perturbation produced by an operator insertion inside the $n$-point functions as it follows \cite{Clark:2004sb,Dibitetto:2017klx}
\begin{equation}
\begin{split}
\langle\mathcal{O}_{1}(x_{1})\cdots&\mathcal{O}_{n}(x_{n})\rangle_{\textrm{def.}}  =  \langle\mathcal{O}_{1}(x_{1})\cdots\mathcal{O}_{n}(x_{n})\rangle_{0} \\
& +  \gamma\,\int \D^{5}z \,\phi(z)\,\langle\mathcal{O}_{1}(x_{1})\cdots\mathcal{O}_{n}(x_{n})\,\mathcal{O}_{X}(z)\rangle_{0}  \, \\
 &+ \frac{\gamma^{2}}{2!}\,\int \D^{5}z\int d^{5}w \,\phi(z)\phi(w)\,\langle\mathcal{O}_{1}(x_{1})\cdots\mathcal{O}_{n}(x_{n})\,\mathcal{O}_{X}(z)\,\mathcal{O}_{X}(w)\rangle_{0} \, + \, \dots 
\end{split}
\end{equation}
If we now choose $\ma O_1=\ma O_X$, we obtain
\begin{equation}
\langle\mathcal{O}_{X}(\rho)\rangle_{\textrm{def.}}\,=\,\underbrace{\langle\mathcal{O}_{X}(\rho)\rangle_{0}}_{0}
\,+\,\gamma\,\int \D^{5}z\,\phi(z)\,\underbrace{\langle\mathcal{O}_{X}(\rho)\mathcal{O}_{X}(z)\rangle_{0}}_{\frac{a}{|\rho-z|^{6}}}\,+\,\dots
\label{1pointfgeneral}
\end{equation}
Performing the integral in \eqref{1pointfgeneral}, it follows that, if $\phi(\rho)\sim\rho^{-2}$, whence the 1-point function associated to $\ma O_X$ is given by
\begin{equation}
  \langle \ma O_X \rangle =\frac{2\,\pi^{2}a\,\gamma}{3\,\rho^3}\ .
  \label{1pointbulk}
\end{equation}
We conclude that the $\rho^{-3}$ dependence of \eqref{1pointbulk} matches non-trivially with the holographic result \eqref{1pointbulk}. As far as a more complete matching is concerned (\emph{i.e.} including the parameters $a,c$ and $\gamma$), it would require a more rigourous derivation considering the explicit form of the parameters inside correlators and the Lagrangian of the $\ma N=2$ $\mrm{SCFT}_5$.


\section*{Acknowledgements}

We would like to thank Y.~Lozano for very interesting and stimulating discussions.
NP would also like to acknowledge the members of theoretical group of IPM, Tehran, for their kind hospitality while part of this work was being prepared. The work of GD is supported by the Swedish Research Council (VR). The work of NP was partially supported by ICTP. 

\appendix

\section{Massive IIA Supergravity}
\label{app:massiveIIA}

In this appendix we review the main features of massive IIA supergravity \cite{Romans:1985tz}. The theory is characterized by the bosonic fields $g_{MN}$, $\Phi$, $B_{(2)}$, $C_{(1)}$ and $C_{(3)}$. The action has the following form
\begin{equation}
  S_{\mathrm{mIIA}}=\frac{1}{2\kappa_{10}^2}\,\biggl [\int \D^{10}x\,\sqrt{-g}\,e^{-2\Phi}\left(R+4\,\partial_{\mu}\,\Phi\,\partial^{\mu}\,\Phi-\frac12\,|H_{(3)}|^2   \right)-\frac12 \sum_{p=0,2,4} |F_{(p)}|^2  \biggr]+S_{\text{top}}\ ,
  \label{massiveIIAaction}
\end{equation}
where $S_{\text{top}}$ is a topological term given by
\begin{equation}
\begin{split}
S_{\text{top}}=&-\frac{1}{2} \int ( B_{(2)}\wedge F_{(4)}\wedge F_{(4)}-\frac13  F_{(0)}\wedge B_{(2)}\wedge B_{(2)}\wedge B_{(2)}\wedge F_{(4)}\\
&+\frac{1}{20}F_{(0)}\wedge F_{(0)}\wedge B_{(2)}\wedge B_{(2)}\wedge B_{(2)}\wedge B_{(2)}\wedge B_{(2)})\ ,
\end{split}
\end{equation}
where $H_{(3)}=dB_{(2)}$, $F_{(2)}=dC_{(1)}$, $F_{(3)}=dC_{(3)}$ and the 0-form field strength $F_{(0)}$ is associated to the Romans' mass as $F_{(0)}=m$.

All the equations of motion can be derived\footnote{We set $\kappa_{10}=8\pi G_{10}=1$.} consistently from \eqref{massiveIIAaction}. They have the following form
\begin{equation}
 \begin{split}
  R_{MN}-\frac12\,T_{MN}&=0\ ,\\[1mm]
  \Box\Phi-|\partial \Phi|^2+\frac14R-\frac18 |H_{(3)}|^2&=0\ ,\\[1mm]
  d\left(e^{-2\Phi}\star_{10}H_{(3)}\right)&=0\ ,\\[1mm]
  \left(d+H_{(3)}\wedge\right)(\star_{10}F_{(p)})&=0\ ,\quad \text{with}\quad p=2,4\ ,
 \end{split}
 \label{massiveIIAeoms}
\end{equation}
where $M, N, \dots=0,\dots, 9$ and $R$ and $\Box$ are respectively the 10d scalar curvature and the Laplacian. The stress-energy tensor is given by
\begin{equation}
\begin{split}
 T_{MN}&=e^{2\Phi}\sum_p \left(\frac{p}{p!}\,F_{(p)\,MM_1\dots M_{p-1}}F_{(p)\,N}^{\qquad M_1\dots M_{p-1}}-\frac{p-1}{8}g_{MN}|F_{(p)}|^2 \right)\\
 &+\left(\frac12\,H_{(3)\,MPQ}H_{(3)\,N}^{\qquad PQ}-\frac14 g_{MN}|H_{(3)}|^2 \right)-\left(4\nabla_M \nabla_N \Phi +\frac12 g_{MN}(\Box \Phi-2|\partial \Phi|^2) \right)\,,
 \end{split}
\end{equation}
with $\nabla_M$ being associated with the Levi-Civita connection of the 10d background. The Bianchi identities take the form
\begin{equation}
\begin{split}
  dF_{(2)}&= F_{(0)}\wedge H_{(3)}\ ,\\
  d F_{(4)}&=-F_{(2)}\wedge H_{(3)}\ ,\\
 d H_{(3)}&=0\ ,\\
  d F_{(0)}&=0\ .
 \label{massiveIIAbianchi}
\end{split}
\end{equation}
As a consequence of \eqref{massiveIIAbianchi}, the following fluxes 
\begin{equation}
 m,\qquad H_{(3)}\,\qquad F_{(2)}-m B_{(2)}\,,\qquad F_{(4)}-B_{(2)}\wedge F_{(2)}+\frac12\,mB_{(2)}\wedge B_{(2)}\,,
\end{equation}
turn out to satisfy a Dirac quantization condition.

It may be worth mentioning that the truncation Ansatz of section \ref{reduction} is obtained by casting massive IIA supergravity into the Einstein frame \cite{Cvetic:1999un}. To convert the action \eqref{massiveIIAaction}, the equations of motions \eqref{massiveIIAeoms} and Bianchi identities \eqref{massiveIIAbianchi} into the Einstein frame, one has to redefine the metric as $g_{MN}=e^{\Phi/2}\,g^{(\mathrm{E})}_{MN}$.

\section{Symplectic-Majorana-Weyl Spinors in $d=1+5$}
\label{app:SM_spinors}

In this appendix we collect the conventions and the fundamental relations involving irreducible spinors in $d=1+5$. Subsequently, we construct an explicit representation of Dirac matrices. In $d=1+5$ Dirac spinors enjoy 16 independent real components and they can be decomposed into irreducible Weyl spinors with opposite chirality and having 8 independent real components each. The 6d Clifford algebra is defined by the relation
\begin{equation}
\left\{\Gamma^{m},\,\Gamma^{n}\right\} = 2\,\eta^{mn} \, \mathbb{I}_{8} \ ,
\label{clifford6d}
\end{equation}
where $\left\{\Gamma^{m}\right\}_{m\,=\,0,\,\cdots\,5}$ are the $8\times 8$ Dirac matrices and $\eta=\text{diag}(-1,+1,+1,+1,+1)$. The chirality operator $\Gamma_*$ can be defined in the following way in terms of the above Dirac matrices 
\begin{equation}
\label{gammastar}
 \Gamma_*=\Gamma^0\,\Gamma^1\,\Gamma^2\,\Gamma^3\,\Gamma^4\,\Gamma^5\qquad \text{with}\qquad \Gamma_*\,\Gamma_*=\mathbb{I}_{8}\ .
\end{equation}
For $(1+5)$-dimensional backgrounds, we can choose the matrices $A, B, C$, respectively realizing Dirac, complex and charge conjugation, satisfying the following defining relations \cite{VanProeyen:1999ni}
\begin{equation}
 \left(\Gamma^{m}\right)^{\dagger}  =  -A \, \Gamma^{m} \, A^{-1} \,,\quad   \left(\Gamma^{m}\right)^{*}  = B \, \Gamma^{m} \, B^{-1}\,, \quad  \left(\Gamma^{m}\right)^{T}  =  -C \, \Gamma^{m} \, C^{-1} \ ,
\end{equation}
with 
\begin{equation}
 B^{T} = C \, A^{-1} \ ,\quad  B^{*}\,B  = -\mathbb{I}_{8} \ ,\quad C^{T}  =  -C^{-1}  =  -C^{\dagger}  =  C \  .
 \label{ABCidentities}
\end{equation}
The second identity in \eqref{ABCidentities} implies that it is actually inconsistent to define a proper reality condition on Dirac (or Weyl) spinors.
However, it is always possible to introduce $\mathrm{SU}(2)_R$ doublets $\zeta^a$ of Dirac spinors, called symplectic-Majorana (SM) spinors respecting a pseudo-reality condition \cite{VanProeyen:1999ni} given by
\begin{equation}
 \label{SM_cond}
\zeta_{a} \equiv\left(\zeta^{a}\right)^{*}  \overset{!}{=}  \epsilon_{ab}\,B\,\zeta^{b} \  ,
\end{equation}
where $\epsilon_{ab}$ is the $\mathrm{SU}(2)$ invariant Levi-Civita symbol. The condition \eqref{SM_cond} ensures us that the number of independent components of a SM spinor be the same of those of a Dirac spinor. Moreover, the above condition also turns out to be compatible with the projections onto the chiral components of a Dirac spinor. Hence it is possible to construct SM doublets of irreducible Weyl spinors that are called symplectic-Majorana-Weyl (SMW) spinors.

Let us now construct an explicit representation for the Dirac matrices satisfying \eqref{clifford6d}. We firstly introduce the Dirac matrices $\left\{\rho^\alpha \right\}_{\alpha\,=\,0,\,1\,,2}$ for a $(1+2)$-dimensional background in the Majorana representation as it follows
\begin{equation}
 \rho^0=i\sigma^2\ ,\qquad  \rho^1=\sigma^1\ ,\qquad  \rho^2=\sigma^3\ ,
 \label{rho3d}
\end{equation}
and the Dirac matrices for a Euclidean 2-dimensional background $\left\{\gamma^i \right\}_{i\,=1\,,2}$ as
\begin{equation}
 \gamma^1=\sigma^1\ ,\qquad \gamma^2=\sigma^3\ ,\qquad \gamma_*=i\gamma^1\gamma^2=\sigma^2\ ,
 \label{gammai}
\end{equation}
where
\begin{equation}
 \label{Pauli}
\sigma^{1} =
\left(
\begin{array}{cc}
0 & 1  \\
1 & 0
\end{array}
\right)
\hspace{5mm} \textrm{ , } \hspace{5mm}
\sigma^{2} =
\left(
\begin{array}{cc}
0 & -i  \\
i & 0
\end{array}
\right)
\hspace{5mm} \textrm{ , } \hspace{5mm}
\sigma^{3} =
\left(
\begin{array}{cc}
1 & 0  \\
0 & -1
\end{array}
\right) \ .
\end{equation}
are the Pauli matrices. An explicit representation of the $(1+5)$-dimensional Dirac matriced satisying \eqref{clifford6d} can be defined in the following way
\begin{equation}
 \begin{split}
  \Gamma^\alpha&=\rho^\alpha \,\otimes \, \mathbb{I}_2 \,\otimes \, \sigma^1 \ ,\qquad \text{with}\qquad \alpha=0,\,1,\,2\ ,\\
   \Gamma^3&=\mathbb{I}_2 \,\otimes \, \gamma_* \,\otimes \, \sigma^2\ ,\\
    \Gamma^i&=\mathbb{I}_2 \,\otimes \, \gamma^i \,\otimes \, \sigma^2\ ,\qquad \,\text{with}\qquad \,i=1,\,2\ .
    \label{gammamatrices}
 \end{split}
\end{equation}
In this representation the chirality operator \eqref{gammastar} takes the form
\begin{equation}
 \Gamma_*=\mathbb{I}_2 \,\otimes \, \mathbb{I}_2 \,\otimes \, \sigma^3\ ,
 \label{gammastarrep}
\end{equation}
while the matrices $A, B, C$ may be written as
\begin{equation}
 \begin{split}
  A&=\Gamma^0=i\,\sigma^2\,\otimes \, \mathbb{I}_2 \,\otimes \, \sigma^1\ ,\\
  B&=-i\,\Gamma^4\,\Gamma^5=-\mathbb{I}_2 \,\otimes \, \gamma_*\,\otimes \, \mathbb{I}_2\  ,\\
  C&=i\,\Gamma^0\,\,\Gamma^4\,\Gamma^5=i\,\sigma^2 \,\otimes \, \gamma_*\,\otimes \, \sigma^1 \ .
  \label{abcrep}
 \end{split}
\end{equation}

\section{Gauged $\mathcal{N}=(1,1)$ Supergravities in Six Dimensions}
\label{app:halfmax}

Half-maximal supergravities in $(1+5)$ spacetime dimensions enjoy sixteen real supercharges. As we have seen in the previous appendix, these can be organized into two chiral spinors.
As a consequence, just as in $(1+9)$ dimensions, we have the choice of picking both spinors with the same (iib), or opposite (iia) chiralities. In this paper we are only interested in the latter case, \emph{i.e.} $\mathcal{N}=(1,1)$ supergravities.
The goal of this appendix is that of giving an overview of consistent embedding tensor deformations of these theories and understanding what particular choice gives rise to the Romans' theory.

If we start from a maximal theory with $\mathcal{N}=(2,2)$ supersymmetry and $\mathrm{SO}(5,5)$ global symmetry, its fields can be rearranged as shown in table~\ref{table:fields_max}.
\begin{table}[h!]
\renewcommand{\arraystretch}{1}
\begin{center}
\scalebox{1}[1]{
\begin{tabular}{c| c | c}
Field Type & Field Name & $\mathrm{SO}(5,5)$ irrep's \\
\hline \hline
Scalars & $\mathcal{V}_{A}^{\phantom{A}\alpha\dot{\alpha}}$ & $\textbf{45}$ ($25$ phys.) \\
\hline
Vectors & $A_{\mu}^{\phantom{A}A}$ & $\textbf{16}_{c}$ \\
\hline
Two-forms & $\mathcal{B}_{\mu\nu}^{\phantom{\mu\nu}M}$ & $\textbf{10}$ ($5$ phys.) \\
\hline
\end{tabular}
}
\end{center}
\caption{{\it The (bosonic) field content of maximal supergravity in six dimensions. $\mu$ denotes a spacetime index, $A$ is a MW spinor of $\mathrm{SO}(5,5)$, $M$ is an $\mathrm{SO}(5,5)$ fundamental index, while $\alpha$ \& $\dot{\alpha}$ denote spinors of the time(space)like $\mathrm{SO}(5)$ subgroups of $\mathrm{SO}(5,5)$.}} \label{table:fields_max}
\end{table}

The embedding tensor deformations were exhaustively studied in \cite{Bergshoeff:2007ef} and can be arranged into a unique $\mathrm{SO}(5,5)$ irrep, \emph{i.e.} $\Theta\in\textbf{144}_{c}$.
Following now the philosophy of \cite{Dibitetto:2011eu}, we identify a $\mathbb{Z}_{2}$ that partially breaks supersymmetry down to $\mathcal{N}=(1,1)$, while retaining the correct field content.
This is realized by
\begin{equation}
\begin{array}{cccccc}
\mathrm{SO}(5,5) & & \overset{\mathbb{Z}_{2}}{\supset} & & \mathbb{R}^{+}\times\mathrm{SO}(4,4)  & , \\[2mm]
\textbf{10} & & \longrightarrow & & \underbrace{\textbf{1}^{(+2)} \,\oplus\,\textbf{1}^{(-2)}}_{\textrm{even}} \,\oplus\,\underbrace{\textbf{8}_{v}^{(0)}}_{\textrm{odd}} & .
\end{array}
\end{equation}
The above $\mathbb{Z}_{2}$ is completely specified by assigning \emph{even} parity to the $\textbf{8}_{c}$ of $\mathrm{SO}(4,4)$, while keeping the $\textbf{8}_{s}$ \& $\textbf{8}_{v}$ parity-\emph{odd}.
Correspondingly, the $\mathrm{SO}(5)^{2}$ R-symmetry group of the maximal theory is broken to $\mathrm{SO}(4)^{2}$, its diagonal $\mathrm{SO}(4)_{\textrm{diag}}=\mathrm{SU}(2)_{\textrm{L}}\times\mathrm{SU}(2)_{\textrm{R}}$ subgroup being the R-symmetry group of the half-maximal theory.
The supercharges branch as
\begin{equation}
\begin{array}{cccccc}
\mathrm{SO}(5)\times\mathrm{SO}(5) & & \overset{\mathbb{Z}_{2}}{\supset} & & \mathrm{SU}(2)_{\textrm{L}}\times\mathrm{SU}(2)_{\textrm{R}}  & , \\[2mm]
(\textbf{4},\textbf{4}) & & \longrightarrow & & \underbrace{(\textbf{2},\textbf{2})}_{\textrm{even}} \,\oplus\,\underbrace{(\textbf{1},\textbf{1}) \,\oplus\,(\textbf{3},\textbf{1})}_{\textrm{odd}} & .
\end{array}
\end{equation}

This procedure gives rise to a half-maximal theory coupled to four vector multiplets, whose field content is summarized in table~\ref{table:fields_halfmax}.
\begin{table}[h!]
\renewcommand{\arraystretch}{1}
\begin{center}
\scalebox{1}[1]{
\begin{tabular}{c| c | c}
Field Type & Field Name & $\mathbb{R}^{+}\times\mathrm{SO}(4,4)$ irrep's \\
\hline \hline
Scalars & $X$ \& $\mathcal{M}_{MN}$ & $\textbf{1}^{(0)}\,\oplus\,\textbf{28}^{(0)}$ ($17$ phys.) \\
\hline
Vectors & $A_{\mu}^{\phantom{A}M}$ & $\textbf{8}^{(+1)}$ \\
\hline
Two-forms & $\mathcal{B}_{\mu\nu}^{\phantom{\mu\nu}\pm}$ & $\textbf{1}^{(+2)}\,\oplus\,\textbf{1}^{(-2)}$ ($1$ phys.) \\
\hline
\end{tabular}
}
\end{center}
\caption{{\it The (bosonic) field content of half-maximal supergravity in six dimensions. $\mu$ denotes a spacetime index, $M$ is a MW spinor of $\mathrm{SO}(4,4)$. Note that the degrees of freedom of the two-form are halved by means of a self-duality condition.}} \label{table:fields_halfmax}
\end{table}

The consistent embedding tensor deformations can be obtained by branching the $\textbf{144}_{c}$ of $\mathrm{SO}(5,5)$ w.r.t. its $\mathbb{R}^{+}\times\mathrm{SO}(4,4)$ subgroup and only retaining the parity-even irrep's.
This yields
\begin{equation}
\begin{array}{lccclc}
\Theta & & \in & & \underbrace{\textbf{8}_{c}^{(-1)}}_{\xi_{M}}\,\oplus\,\underbrace{\textbf{8}_{c}^{(+3)}}_{\zeta_{M}}\,\oplus\,\underbrace{\textbf{56}_{c}^{(-1)}}_{f_{[MNP]}} & ,
\end{array}
\end{equation}
where $\xi$ \& $f$ parametrize gaugings, of $\mathbb{R}^{+}$ and $\mathrm{SO}(4,4)$ respectively, while $\zeta$ represents a massive deformation inducing a St\"uckelberg coupling for the two-form.
Such embedding tensor needs to satisfy the following quadratic constraint (QC) for consistency (when $\xi_{M}=0$)
\begin{equation}
\begin{array}{lcccclc}
f_{R[MN}\,f_{PQ]}^{\phantom{PQ]}R} \ = \ 0 & , & & & & f_{MNP}\,\zeta^{P} \ = \ 0 & ,
\end{array}\label{QC}
\end{equation}
where all contractions are taken w.r.t.~the $\mathrm{SO}(4,4)$ invariant metric $\eta_{MN}$. 
It is worth mentioning that, in order for our half-maximal theory to still admit an embedding within the maximal theory, the following set of \emph{extra} QC is required
\begin{equation}
\begin{array}{lcccclc}
f_{MNP}\,f^{MNP} \ = \ 0 & , & & & & \left.f_{[MNP}\,\zeta_{Q]}\right|_{\textrm{SD}} \ = \ 0 & ,
\end{array}\label{extraQC}
\end{equation}
where $|_{\textrm{SD}}$ denotes the projection on the self-dual four-form of $\mathrm{SO}(4,4)$.

The scalar potential induced by the above deformations (after setting $\xi_{M}=0$) reads
\begin{equation}
\begin{split}
V=&f_{MNP}f_{QRS}X^{2}\left(\frac{1}{12}\mathcal{M}^{MQ}\mathcal{M}^{NR}\mathcal{M}^{PS}-\frac{1}{4}\mathcal{M}^{MQ}\eta^{NR}\eta^{PS}+\frac{1}{6}\eta^{MQ}\eta^{NR}\eta^{PS}\right)\\
&+\frac{1}{2}\zeta_{M}\zeta_{N}X^{-6}\mathcal{M}^{MN}\,+\,\frac{2}{3}f_{MNP}\zeta_{Q}X^{-2}\mathcal{M}^{MNPQ} \ ,
\end{split}\label{Half_Max_Pot}
\end{equation}
where $\mathcal{M}^{MNPQ}=\frac{1}{4!}\epsilon_{abcd}\mathcal{V}^{Ma}\mathcal{V}^{Nb}\mathcal{V}^{Pc}\mathcal{V}^{Qd}$, $\mathcal{V}$ being the ``vielbein'' reproducing the scalar coset representative $\mathcal{M}$.

The Romans' theory is further obtained by truncating away \emph{all} scalars but $X$, \emph{i.e.} picking $\mathcal{M}_{MN}=\delta_{MN}$, and keeping only fields with legs within the timelike $\mathrm{SO}(4)$, call it $a=0,\,i$, where $i=1,2,3$.
Finally pick the following embedding tensor
\begin{equation}
\begin{array}{lccclc}
f_{ijk} \ = \ g\,\epsilon_{ijk} & , & & & \zeta_{0} \ = \ -\sqrt{2}\,m & ,
\end{array}
\label{ET_Romans} 
\end{equation}
all the other components being zero. This choice can be checked to satisfy the QC in \eqref{QC}, which are needed for consistency. 
The scalar potential \eqref{Half_Max_Pot} specified to this case reads
\begin{equation}
V(X)\,=\,m^2\,X^{-6}-4\sqrt{2}\,gm\,X^{-2}-2\,g^2\,X^2\ ,
\end{equation}
which precisely reproduces \eqref{scalarpotential}.

As a final comment, we note that the \emph{extra} QC \eqref{extraQC} needed for a consistent embedding in maximal supergravity are actually violated by \eqref{ET_Romans}, thus suggesting the presence of spacetime-filling branes within the massive IIA realization of this theory.

 \bibliographystyle{utphys}
  \bibliography{references}
\end{document}